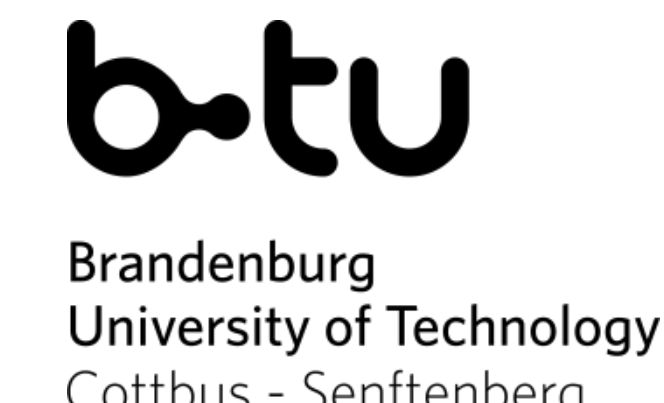

# Working Paper

## Adaptive Robust Optimization for European Electricity System Planning Considering Regional Dunkelflaute Events

* Maximilian Bernecker (BTU Cottbus-Senftenberg), maximilian.bernecker@b-tu.de

Smaranda Sgarciu (BTU Cottbus-Senftenberg), smaranda.sgarciu@b-tu.de

Xiaoming Kan (Stockholm University) xiaoming.kan@dsv.su.se

Mehrnaz Anvari (Fraunhofer) (Fraunhofer SCAI), mehrnaz.anvari@scai.fraunhofer.de

Iegor Riepin (TU Berlin), iegor.riepin@tu-berlin.de

Felix Müsgens (BTU Cottbus-Senftenberg), felix.muesgens@b-tu.de

* Corresponding author.

*E-mail address:* maximilian.bernecker@b-tu.de (M. Bernecker)

**Chair of Energy Economics**
**Brandenburg University of Technology Cottbus-Senftenberg, Cottbus, Germany**
Prof. Dr. Felix Müsgens
http://www.b-tu.de/fg-energiewirtschaft/

# Abstract

The expansion of wind and solar power is driving the European energy system transformation, thereby also driving our reliance on this weather-dependent resources. Integrating renewable scarcity events into long-term planning has therefore become essential. This study demonstrates how worst-case regional renewable scarcity events - such as the Dunkelflaute, prolonged periods of low wind and solar availability - can be incorporated endogenously into the planning of a weather-robust, interconnected energy system. We develop a capacity expansion model for a fully decarbonized European electricity system using an adaptive robust optimization framework that incorporates multiple extreme weather realizations within a single optimization run. Results show that system costs rise nonlinearly with the geographic extent of these events: a single worst-case regional disruption increases costs by 9%, but broader disruptions across multiple regions lead to much sharper increases, up to 51%. As *Dunkelflaute* conditions extend across most of Europe, additional cost impacts level off, with a maximum increase of 71%. The optimal technology mix evolves with the severity of weather stress: while renewables, batteries, and interregional transmission are sufficient to manage localized events, large-scale disruptions require long-term hydrogen storage and load shedding to maintain system resilience. Central European regions, especially Germany and France, emerge as systemic bottlenecks, while peripheral regions bear the cost of compensatory overbuilding. These findings underscore the need for a coordinated European policy strategy that goes beyond national planning to support cross-border infrastructure investment, scale up flexible technologies such as long-duration storage, and promote a geographically balanced deployment of renewables to mitigate systemic risks associated with Dunkelflaute events.

# 1 INTRODUCTION

Limiting the increase in global average temperature to "well below" 2 °C requires nearly zero or even negative $CO_2$ emissions by mid-21$^{st}$ century (Rogelj et al., 2015; Davis et al., 2018). Variable renewable energy technologies such as wind power and solar photovoltaic (PV) are expected to form the foundation of future low-carbon electricity system (Schlachtberger et al., 2017; Reichenberg et al., 2018; Brown et al., 2018; Mattsson et al., 2021). As low-carbon electricity systems based on variable renewable energy are highly weather-dependent, weather-induced uncertainties, particularly the risk of prolonged periods of low renewable output, have become a central focus of both political discussions and academic research. The term "Dunkelflaute" has become popular, even in the English language, to describe periods when the availability of wind and solar power is low, particularly the non-windy and non-sunny periods that manifest for several consecutive days (Li et al., 2020, 2021a, 2021b; Mayer et al., 2023).

It is crucial that future energy systems are designed to withstand such fluctuations in renewable resource availability. The development of mathematical optimization programs for energy system planning introduces various methodological approaches to address this challenge. Historically, most approaches exploring optimal energy systems were deterministic, based on expected or "typical" weather years. To ensure a certain level of resilience, such an optimization can be also based on a typical "worst-case" weather year. Alternatively, sampling over several historical weather years offers heuristic approaches to achieve resilience, such as averaging over the samples or imposing upper and lower bounds on generation capacities (Caglayan et al., 2021; Gøtske et al., 2024). However, a drawback is that deterministic optimization can analyze uncertainty in a post processing step only, based on exogenous parameter changes. This omits key trade-offs in decision-making, such as the coupling between proactive investments made in anticipation of uncertainty and reactive measures taken once uncertainties materialize. Furthermore, a key difficulty not addressed in these approaches, is that weather events causing extended periods of renewable scarcity, so called *Dunkelflaute,* vary across Europe in their frequency, intensity, and timing, and often affect countries unevenly. In the following we refer to the term "renewable scarcity periods" as periods of low electricity production potential from wind and/ or solar generation resources. In addition, we use the term "Dunkelflaute" to refer to a period of low electricity production potential from wind and solar resources occurring simultaneously within the same geographical zone. The impacts of such Dunkelflaute can ripple across borders, disrupt electricity flows and threatening system stability well beyond the regions directly affected.

Advances in hardware and software have allowed an expanding area of energy-system planning research to model stochasticity endogenously. There exist several established approaches for modeling uncertainty in energy optimization problems. A comprehensive overview of uncertainty modeling frameworks in energy system optimization is provided by Roald et al. ( 2023). Stochastic optimization is one of the most widely used approaches. Stochastic optimization represents uncertainty using probabilistic scenarios and optimizes expected system costs (Birge and Louveaux, 2011). While it enables recourse decisions, it requires reliable probability distributions and a sufficiently large number of scenarios to accurately represent uncertain parameters. This often leads to computational scaling issues, as model size increases rapidly with the number of scenarios, making large-scale energy system applications computationally challenging (Conejo et al. 2016). Applying stochastic optimization at the European scale to capture renewable scarcity events for robust long-term planning is particularly challenging because uncertainty is high-dimensional in both space and time. Extreme low-wind and low-solar events may occur in different regions with varying spatial extents, making it difficult to construct a reliable scenario tree.

In contrast, robust optimization, seeks solutions that remain feasible and cost-efficient under the worst possible realization of uncertainties within a prescribed uncertainty set and does not rely on probability distributions (Ben-Tal et al., 2009). Classical static robust optimization formulations, however, can be overly conservative because they do not allow decisions to adapt to the realization of uncertainty. In the context of weather uncertainty, this means that the system is designed to withstand a rare extreme event, which may never actually occur. Adaptive (or adjustable) robust optimization (ARO) introduces decision rules that enable recourse actions and thereby reduce conservatism while retaining computational tractability through standard approximations such as affine decision rules or column-and-constraint generation (Bertsimas et al. 2013). It is particularly well-suited for energy systems exposed to weather-driven uncertainty, as it enables long term planning to incorporate a wide range of uncertainty realizations. In this context, ARO iteratively identifies the worst-case weather realizations and optimizes the system accordingly. Beyond its methodological suitability for the present problem, ARO can also provide a tractable approximation to otherwise intractable multistage stochastic programs with recourse. In particular, under mild assumptions and with increasing data, robust solutions obtained via decision rules can approximate the cost and first-stage decisions of corresponding two-stage stochastic programs, with the robust cost converging to the stochastic optimal cost (Bertsimas et al. 2020).

To the best of our knowledge, our work is the first that endogenously incorporates worst-case renewable scarcity events into a European electricity system model using the ARO method. We apply this theoretical framework to design a decarbonized electricity system for Europe in 2050 that is robust against periods of extremely low wind and solar availability. In contrast to existing studies that rely on deterministic simulations of historical weather years, our model is based on a specifically defined uncertainty set derived from historical weather data. This formulation enables the generation of multiple synthetic extreme scenarios within a single optimization run and allows tracing the origin of critical worst-case realizations and their influence on system adaptation decisions, ultimately yielding a weather-robust capacity mix. Furthermore, our work advances the literature by incorporating a regional perspective into the analysis of capacity adaptation mechanisms, allowing us to quantify how the burdens of system adaptations are distributed within Europe. To this end, we model a whole year and represent low wind and solar availability through geographically defined events across six distinct weather zones in Europe, during a four-week period in January. This enables the model to endogenously determine where and when worst-case renewable scarcity events occur and to proactively prepare for them. The contribution of this manuscript can be summarized as follows:

1. Development an ARO methodology for designing a robust European electricity system that endogenously accounts for regional renewable worst-case renewable scarcity periods such as Dunkelflaute.
2. Identify the region-specific worst-case events across Europe and quantify on both regional and system-wide costs and infrastructure requirements, providing insights into spatial vulnerabilities and adaptation needs in a fully decarbonized power system.

Our study advances the discussion on energy system robustness by deriving a single optimal solution for a European power system that is resilient to multiple endogenously generated weather uncertainty scenarios. Furthermore, we shed light on the system's adaptation mechanisms - that is, the interplay between different power technology expansion options that enhance robustness against a range of extreme renewable scarcity events. Compared to existing studies in the field, we investigate in detail how individual scarcity events are driving the system adaptation. Consequently, this work is interesting for the energy system modeling community and real-world energy system planners.

Section 2 reviews existing literature on weather-related events and methodological approaches to incorporate the resulting uncertainty into energy system planning. We focus particularly on optimization frameworks and empirical energy system models, thereby revealing the remaining research gap. In section 3, we introduce the underlying ARO model. Our case study is presented in section 4, and the results are discussed in section 5. We finish the paper with conclusions and discussion of the key takeaways in section 6.

# 2 LITERATURE REVIEW

Long-term energy system planning with a focus on extreme weather events has become increasingly popular over the last two decades. The majority[1] of papers explore weather events which impact supply and demand. Examples are heatwaves and cold periods, hydro droughts, and low solar and wind availability periods. All these topics have become relevant for electricity system planning. Low-output periods of intermittent renewables are receiving increasing attention both within academia and more broadly[2]. For the European energy system, the impact of reduced RES availability is of major interest since a 100 % renewable European system in 2050 will heavily rely on wind and solar generation technologies. Therefore, several academic studies investigate such weather events and their impacts, as well as the requirements on the current and future electricity systems to guarantee a reliable energy supply.

These investigations can be distinguished as (1) meteorological approaches, describing historical extreme weather events (e.g. Ohlendorf and Schill, 2020; Li et al., 2020, 2021a, 2021b; Jurasz et al., 2021), (2) theoretical optimization approaches, focusing on weather-related uncertainties in the planning of test systems (e.g. Jabr, 2013; Roldán et al., 2018; Roldán et al., 2019; Roldán et al., 2020) and (3) empirical energy system planning models that incorporate meteorological data to address weather related uncertainty (Perera et al., 2020; Ruhnau and Qvist, 2022; Plaga and Bertsch, 2022; Grochowicz et al., 2023, 2024; Gøtske et al., 2024). As our study focusses on modeling extreme weather events within an energy system optimization framework, we place particular emphasis on reviewing the latter two categories.

Our review begins with category (2), which covers studies that address theoretical optimization approaches. In 2013, Jabr designed a robust optimization model to include the uncertainty from renewable generation and electricity load in the transmission network expansion planning process using a cardinality-constrained uncertainty set. The developed bi-level optimization problem was iteratively solved using Bender's decomposition approach (an iterative technique for mathematical programming). Testing and comparing the proposed model on various test grid systems showed that the method consistently generated reliable and cost-effective expansion plans. In a series of follow-up papers, Roldán et al., (2018, 2019, 2020), present robust model formulations, to tackle the incorporation of weather uncertainties into the transmission and generation expansion planning problem. A three-level adaptive robust optimization program was introduced to include uncertainty from wind turbines in a dynamic investment (Roldán et al., 2018). To address the disregarded correlation of uncertainty sources such as wind speed, Roldán et al. (2019) turn from using cardinal and polyhedral sets towards so-called ellipsoidal uncertainty sets. The newly developed two-stage adaptive robust optimization approach was developed further by Roldán et al. (2020) to consider the complete probability structure of uncertainty parameters, ensuring that the optimal solution found is valid. Baringo et al. (2020) developed an adaptive robust optimization model to address the expansion planning of a distribution system that incorporates solar powered electric vehicles and energy storage systems. The model considers both short-term variabilities, such as daily demand patterns, renewable variation, and long-term uncertainty, such as future peak demand and the penetration of electric vehicles. The adaptive robust optimization methodology proves to be an effective tool for distribution expansion planning by highlighting that incorporating long-term uncertainty significantly influences investment decisions. However, all these studies are conducted on test grid systems as the focus is on validating the efficiency of the optimization approaches. Consequently, the transfer of implications on real-world energy systems is limited.

[1] A smaller part explores how extreme events such as floods and (ice) storms impair energy infrastructure.

[2] Some examples: Quartz: https://qz.com/can-europe-survive-the-dreaded-dunkelflaute-1849886529, a children's book named "Dunkelflaute: A Story of Renewable Wind Power" by Stephanie O'Connor (2022), , Madra Rua Publishing, Donegal, Ireland, Reuters: https://www.reuters.com/markets/commodities/german-wind-reliant-power-firms-brace-annual-dunkelflaute-2024-02-21/

These aspects are addressed in empirical studies which integrate weather-related uncertainty into energy system planning models (3). We will review that part of the literature - particularly in the European context - in the following sections to identify and synthesize the existing research gap. Perera et al. (2020) developed a hybrid stochastic-robust optimization method to investigate the impact of extreme weather events using 13 climate change scenarios on 30 cities in Sweden. They state that future climatic fluctuations and uncertainties in renewable energy potential must be incorporated into the energy system planning using suitable methods to incorporate this uncertainty endogenously. Failing to do so could hinder the integration of renewable energy into future systems and compromise the reliability of energy supply. However, the focus on Swedish cities limits the broader applicability of the findings to the European system.

In 2021, Caglayan et al. proposed a methodology to design a 100 % renewable European energy system, including hydrogen infrastructure with the aim of ensuring resilience across 38 historical weather years Their approach relies on a sequential deterministic optimization framework, and their results indicate that incorporating multiple weather years can increase system costs by approximately 25 %. However, the robust system layout is restricted by several heuristic simplifications Notably, the capacity layout for all technologies except biomass is determined based on 30 typical days from a single weather year. This limits the model's capability to incorporate rare extreme events, which are decisive for robust planning in renewable dominated systems. In a subsequent step, the model is run at hourly resolution with these capacities fixed, leaving biomass as the only technology allowed to adjust freely. This overestimates the role of biomass and underestimates the role of other flexible assets, such as long-term hydrogen, resulting in suboptimal solutions.

By employing a deterministic multi-year energy system optimization model, Ruhnau and Qvist (2022) investigated low renewable availability events within the context of a German market that would be fully powered by renewable energy systems in 2050. They analyze a period of 35 historical years, highlighting that periods of low renewable availability can continue for up to two weeks duration. They also point out that the biggest energy shortfall can cover a substantially longer period of up to nine weeks. The authors argue that focusing on short duration renewable scarcity periods or single years may result in underestimating both the storage requirements and costs of a system supplied solely by renewable sources. A limitation of the study is its focus on Germany, which is modeled deterministically as a copper plate with fixed cross-border electricity flows, thereby neglecting the broader European system perspective.

Plaga and Bertsch (2022) apply a deterministic optimization approach to European electricity system planning under climate uncertainty. They analyze six climate scenarios from the EURO-CORDEX database individually and compare them to configurations in which multiple scenarios are modeled sequentially within a single optimization run. The results show that the system cost varies between the cheapest and most expensive scenarios by around 24 percent. Furthermore, the results show that more generation and battery storage capacity increases resilience. While their analysis focuses on the European energy system, the study is limited by analyzing only six different weather scenarios. Furthermore, a robust optimization approach that endogenously accounts for uncertainty realizations was done in this analysis.

Grochowicz et al. (2023) employed a near-optimal space method to generate a resilient European power system. Their geometric concept involves mapping the individual near-optimal feasible spaces from single deterministic weather-year optimizations, and selecting a single solution at the geometric center, providing a certain tolerance towards infeasibility across all considered periods. They found that robust layouts tend to invest more in onshore wind and solar power, leading to a 50 % $CO_2$ emissions reduction compared to a cost optimal solution. As the method is based on intersecting multiple near-optimal solution spaces to identify cost-effective additional investments that enhance system robustness across various weather realizations, no robust optimization approach is applied. Consequently, no endogenous weather uncertainty realizations were modelled.

Also focusing on the European electricity system, Grochowicz et al. (2024) investigated "system defining events"- defined as periods that significantly impact electricity system costs - across a 41-year timespan. In addition to meteorological analysis, they employed a deterministic energy system model using nodal shadow prices and load shedding as indicators of system stress. The single year optimizations indicate that most events occur between November and February with an average duration of 7 days. However, as the focus of the study is to identify factors that characterize difficult weather events such as low temperatures, low wind, transmission congestion and storage constraints, no robust planning model was applied to incorporate weather uncertainty.

Gøtske et al. (2024) assess the impact of 62 historical weather years on a future sector-coupled, decarbonized European energy system using the PyPSA-Eur model. They conduct individual capacity optimizations for each weather year and then test the robustness of the resulting layouts by performing dispatch optimizations across all other years. Their findings highlight that low wind availability during winter poses the greatest challenge to system stability. However, since the study relies on a deterministic modeling approach, the robust system layout is effectively tailored to a single worst-case weather year. This limits the analysis' ability to evaluate trade-offs between capacity expansion and the broader spectrum of possible weather events not captured by that specific year.

Based on this review, we can conclude the following gaps in the literature:

1. Existing theoretical studies apply robust optimization approaches on test systems, offering valuable theoretical insights but limited empirical contributions.
2. Empirical energy system planning studies rely on deterministic and heuristic optimization approaches rather than robust optimization to incorporate weather-related uncertainty.
3. Existing energy system planning studies focus their analysis primarily on the overall system design, omitting deeper insights into how renewable scarcity events impact regional electricity supply and adaptation strategies.

While the reviewed contributions offer valuable insights, most studies focus on analyzing the effect of extreme events at system level. Furthermore, the exogenous treatment of uncertain renewable availability periods based on historical weather realizations limits the investigation space of future system adaptation needs to specific recorded events. In contrast, we developed a novel uncertainty set within an ARO framework, that captures regional and technology-specific renewable scarcity events derived from historical weather data. Thus, our ARO approach just controls the geographical scope and time dimension of extreme scarcity events. This enables the model to endogenously pinpoint which conditions constitute worst-case scarcity events, enabling us to trace the origin of individual worst-case scenarios and analyze their impact on the results. On this basis we can identify the events that drive system adaptation and associated costs. Despite its strong theoretical foundation, ARO has not yet been applied to address weather-related uncertainty in long-term electricity system planning at a European scale. Integrating regionally differentiated wind and/or solar scarcity periods into a detailed planning model for a decarbonized European power system in 2050 complements existing literature by providing insights into the adaptation process with an analysis down to a regional (country) level. In doing so, we also demonstrate both the practical applicability and the added value of robust optimization approaches for real-world energy system design under deep uncertainty.

# 3 METHODOLOGY

In this study, we formulate a long-term generation and transmission expansion planning problem from a central planner's perspective. The energy system model optimizes investment decisions under uncertainty using the ARO method. The process is explained in more detail in Appendix A. In short, a Master Problem (I) and the Subproblem (II) are established, which can be iteratively solved using the column and constraint generation algorithm. The interested reader is referred to the book by Conejo et al. (2016), which details an extensive explanation of the column and constraint generation algorithm.

## 3.1 Expansion decisions (Master problem)

**Nomenclature**

**Indices**

| | |
|---|---|
| t | Time steps |
| n | Nodes |
| l | Transmission lines, AC and DC |
| r | Solar & wind generation units |
| $\overline{cf}_{r,t}$ | Expected solar & wind capacity factors |
| $\widehat{cf}_{r,t}$ | Solar, wind capacity factors maximum deviation |
| $\widetilde{cf}^{M}_{r,t}$ | Solar, wind capacity factors realization |
| $csf_{psp}$ | Capacity scaling factor for pumping storage plants |

c Conventional (conv.) generation units

b Battery storage unit

h Hydrogen storage unit

ror Hydro run of river plant

rsv Hydro reservoir plant

psp Hydro pumping storage plant

g Geographical weather region

**Sets**

$r(l)$ *Receiving node of transmission line*

$s(l)$ *Sending node of transmission line*

$AC(l)$ *Alternating current Transmission line*

$DC(l)$ *Direct current Transmission line*

$\Phi^D$ Primal variables in Master Problem

$\Phi^U$ Uncertainty set

$\Phi^S$ Dual variables in Subproblem

$\Phi_n^R$ Solar, wind power plants located at node n

$\Phi_n^P$ Conventional power plants located at node n

$\Phi_n^B$ Battery storage units located at node n

$\Phi_n^H$ Hydrogen storage units located at node

$\Phi_n^{ROR}$ Run of river plants located at node n

$\Phi_n^{PSP}$ Pumping storage plants located at node n

$\Phi_n^{RSV}$ Hydro reservoir plant located at node n

$pv(r)$ Solar PV power plants

$wind(r)$ Wind turbines

$ref(n)$ Slack or reference node

**Scalars**

$f^{L1}$ Percentage of flexible demand level 1

$f^{L2}$ Percentage of flexible demand level 2

$f^{L3}$ Percentage of flexible demand level 3

$M$ Large scalar value

$\Gamma_r$ Uncertainty budget for renewable unit

**Parameters**

$dem_{n,t}$ Electricity demand at node and time step [MWh]

$ac_r$ Annualized investments costs for solar & wind generation units [EUR/MW]

$ac_b^{INV}$ Annualized investments costs for battery inverter of battery storage units [EUR/MW]

$ac_b^{STOR}$ Annualized investments costs for battery storage units [EUR/MWh]

$ac_h^{OCGT}$ Annualized investments costs for $H_2$-fired open cycle gas turbine [EUR/MW]

$ac_h^{EL}$ Annualized investments costs for electrolyzer [EUR/MW]

$ac_h^{STOR}$ Annualized investments costs for $H_2$ storage [EUR/MWh]

$ac_l$ Annualized investment costs for transmission line [EUR/MW]

$vc_c$ Variable costs of conv. generation [EUR/MWh]

$sc_n^{LS1}$ Level 1 load shedding costs at node n [EUR/MW]

$sc_n^{LS2}$ Level 2 load shedding costs at node n [EUR/MW]

$sc_n^{LS3}$ Level 3 load shedding costs at node n [EUR/MW]

$\overline{cap}_c$ Existing conv. generation capacity [MW]

$\overline{cap}_{ror}$ Existing ror generation capacity [MW]

$\overline{cap}_{rsv}$ Existing rsv generation capacity [MW]

$af_{rsv,t},$ Hydro reservoir availability factor

$\sigma_{psp}$ Efficiency of pumping storage plant

$\sigma_b$ Efficiency of battery inverter

$\sigma_h^{EL}$ Efficiency of electrolyzer

$\sigma_h^{OCGT}$ Efficiency of $H_2$-fired open cycle gas turbine

$\overline{\text{sub}}_r^{REN}$ Solar, wind capacity in subproblem [MW]

$\overline{\text{sub}}_b^{INV}$ Capacity of battery inverter in subproblem [MW]

$\overline{\text{sub}}_b^{STOR}$ Battery storage capacity in subproblem [MWh]

$\overline{\text{sub}}_h^{OCGT}$ Capacity of $H_2$-fired OCGT in subproblem [MW]

$\overline{\text{sub}}_h^{EL}$ Capacity of electrolyzer in subproblem [MW]

$\overline{\text{sub}}_h^{STOR}$ Hydrogen storage capacity in subproblem [MWh]

$\overline{\text{sub}}_l^{Line}$ Transmission line capacity in subproblem [MW]

LB Lower bound in column constraint algorithm

UB Upper bound in column constraint algorithm

**Positive Variables**

$CAP_r$ Capacity of solar, wind generation unit [MW]

$CAP_b^{INV}$ Capacity of battery inverter unit [MW]

$CAP_b^{STOR}$ Storage capacity of battery unit [MWh]

$CAP_h^{OCGT}$ Capacity of $H_2$-fired OCGT unit [MW]

$CAP_h^{EL}$ Capacity of electrolyzer unit [MW]

$CAP_h^{STOR}$ Storage capacity of hydrogen storage unit [MWh]

$CAP_l$ Transmission line capacity [MW]

$GEN_{c,t}$ Generation of conv. unit [MWh]

$GEN_{r,t}$ Generation of solar & wind unit [MWh]

$GEN_{ror,t}$ Generation of ror. unit [MWh]

$GEN_{rsv,t}$ Generation of rsv. unit [MWh]

$GEN_{psp,t}$ Generation of psp unit [MWh]

$CH_{psp,t}$ Charging of psp. unit [MWh]

$CH_{h,t}$ Charging of hydrogen storage [MWh]

$CH_{b,t}$ Charging of battery unit [MWh]

$LVL_{psp,t}$ Storage filling level of psp. unit [MWh]

$LVL_{b,t}$ Storage filling level of battery storage [MWh]

$LVL_{h,t}$ Storage filling level of hydrogen storage [MWh]

$LS_{n,t}^{L1}$ Load shedding level 1 [MWh]

$LS_{n,t}^{L2}$ Load shedding level 2 [MWh]

$LS_{n,t}^{L2}$ Load shedding level 3 [MWh]

$\widetilde{CF}_{r,t}$ Uncertain renewable capacity factor

$Objective^{Master}$ Objective value master problem

$Objective^{SUB}$ Objective value subproblem

**Free Variables**

$PF_{l,t}$ Line power flow [MWh]

$\theta_{s/r(l)}$ Voltage angle receiving/ sending line

**Binary Variables**

$Z_{r,t}$ Variable representing the deviation from reference solar, wind capacity factor

**Auxiliary Variables**

$\phi_{r,t}^{aux}$ Variable used for linearization

**Dual Variables**

| | | | |
|---|---|---|---|
| $\overline{cap}_{psp}$ | Existing psp generation capacity [MW] | $(\cdot)$ | Dual variables are provided after the corresponding equalities or inequalities separated by a colon |
| $\overline{cap}_{l}$ | Existing transmission line capacity [MW] | | |

The first problem, the Master Problem, is a deterministic linear optimization program utilizing the DC optimal power flow method to minimize the investment and generation costs. A detailed model formulation is provided in the following with the optimization variables are denoted in the set $\Phi^{D}$ ={ $CAP_{pv}; CAP_{wind}$ ; $CAP_{b}^{INV}$ ; $CAP_{b}^{STOR}; CAP_{h}^{OCGT}; CAP_{h}^{EL}$ ; $CAP_{h}^{STOR}; CAP_{l}; GEN_{c,t}; GEN_{r,t}; GEN_{ror,t}; GEN_{rsv,t}; GEN_{psp,t}; CH_{psp,t}; CH_{h,t}; CH_{b,t}; LVL_{psp,t}; LVL_{b,t}; LVL_{h,t}; PF_{l,t}; S_{n,t}^{LS1}$; $S_{n,t}^{LS2}; S_{n,t}^{LS3}; \theta_{s/r(l)}$}:

$$\underset{\Phi^{D}}{Min}: Objective^{Master} = \sum_{pv} CAP_{pv} \cdot ac_{pv} + \sum_{wind} CAP_{wind} \cdot ac_{wind} + \sum_{b} CAP_{b}^{INV} \cdot ac_{b}^{INV} + \sum_{b} CAP_{b}^{STOR} \cdot ac_{b}^{STOR} + \sum_{h} CAP_{h}^{OCGT} \cdot ac_{h}^{OCGT} + \sum_{h} CAP_{h}^{EL} \cdot ac_{h}^{EL} + \sum_{h} CAP_{h}^{STOR} \cdot ac_{h}^{STOR} + \sum_{l} CAP_{l} \cdot ac_{l} + \sum_{c,t} GEN_{c,t} \cdot vc_{c} + \sum_{n,t} LS_{n,t}^{L1} \cdot sc_{n}^{LS1} + \sum_{n,t} LS_{n,t}^{L2} \cdot sc_{n}^{LS2} + \sum_{n,t} LS_{n,t}^{L3} \cdot sc_{n}^{LS3} \tag{1}$$

subject to

$$dem_{n,t}\text{-}LS_{n,t}^{L1}\text{-}LS_{n,t}^{L2}\text{-}LS_{n,t}^{L3} = \sum_{r\in\Phi_{n}^{pv}} GEN_{pv,t} + \sum_{r\in\Phi_{n}^{wInd}} GEN_{wind,t} + \sum_{c\in\Phi_{n}^{c}} GEN_{c,t} + \sum_{ror\in\Phi_{n}^{ROR}} GEN_{ror,t} + \sum_{rsv\in\Phi_{n}^{RSV}} GEN_{rsv,t} + \sum_{psp\in\Phi_{n}^{PSP}} \big(GEN_{ps}\text{-}CH_{psp,t}\big) + \sum_{h\in\Phi_{n}^{H}} \big(GEN_{h,t}\text{-}CH_{h,t}\big) + \sum_{b\in\Phi_{n}^{B}} (GEN_{b,t}\text{-}CH_{b,t}) + \sum_{l\in r(l)} PF_{l,t} \text{-} \sum_{l\in s(l)} PF_{l,t} \ : \lambda_{n,t}, \forall n,t \tag{2}$$

$$GEN_{pv,t} \leq CAP_{pv} \cdot \widetilde{cf}_{pv,t}^{M} : \mu_{pv,t}^{R}, \forall\ pv(r), g, t \tag{3}$$

$$GEN_{wind,t} \leq CAP_{wind} \cdot \widetilde{cf}_{wind,t}^{M} : \mu_{wind,t}^{R}, \forall\ wind(r), g, t \tag{4}$$

$$GEN_{c,t} \leq \overline{cap}_{c} \ : \mu_{c,t}^{Conv}, \forall c, t \tag{5}$$

$$GEN_{ror,t} \leq \overline{cap}_{ror} \ : \mu_{ror,t}^{ROR}, \forall ror, t \tag{6}$$

$$GEN_{rsv,t} \leq \overline{cap}_{rsv} \cdot af_{rsv,t} : \mu_{rsv,t}^{RSV}, \forall rsv, t \tag{7}$$

$$GEN_{psp,t} \leq \overline{cap}_{psp} : \mu_{psp,t}^{DIS}, \forall psp, t \tag{8}$$

$$CH_{psp,t} \leq \overline{cap}_{psp} : \mu_{psp,t}^{CH}, \forall psp, t \tag{9}$$

$$LVL_{psp,t} \leq \overline{cap}_{psp} \cdot csf_{psp} : \mu_{PSP,t}^{CAP}, \forall psp, t \tag{10}$$

$$LVL_{psp,t=1} = \frac{(cap_{psp} \cdot csf_{psp})}{2} + (CH_{psp,t=1} \cdot \sigma_{psp})\text{-}GEN_{psp,t=1} \ : \phi_{PSP,t=1}^{LVL}, \forall psp \tag{11}$$

$$LVL_{psp,t} = LVL_{psp,t\text{-}1} + (CH_{psp,t} \cdot \sigma_{psp})\text{-}GEN_{psp,t} \ : \phi_{PSP,t}^{LVL}, \forall psp, \forall t>1, \tag{12}$$

$$GEN_{b,t} \leq CAP_{b}^{INV} : \mu_{b,t}^{DB}, \forall b, t \tag{13}$$

$$CH_{b,t} \leq CAP_{b}^{INV} : \mu_{b,t}^{CB}, \forall b, t \tag{14}$$

$$LVL_{b,t} \leq CAP_{b}^{STOR} \ : \mu_{b,t}^{CAPB}, \forall b, t \tag{15}$$

$$LVL_{b,t=1} = (CH_{b,t=1} \cdot \sigma_{b})\text{-}GEN_{b,t=1} \ : \phi_{b,t=1}^{LVLB}, \forall b \tag{16}$$

$$LVL_{b,t} = LVL_{b,t\text{-}1} + (CH_{b,t} \cdot \sigma_{b})\text{-}GEN_{b,t} \ : \phi_{b,t}^{LVLB}, \forall b, \forall t>1, \tag{17}$$

$$GEN_{h,t} \leq CAP_{h}^{OCGT} : \mu_{h,t}^{OCGT}, \forall h, t \tag{18}$$

$$CH_{h,t} \leq CAP_{h}^{EL} : \mu_{h,t}^{EL}, \forall h, t \tag{19}$$

$$LVL_{h,t} \leq CAP_h^{STOR} : \mu_{h,t}^{CAPH}, \forall h,t \tag{20}$$

$$LVL_{h,t=1} = (CH_{h,t=1} \cdot \sigma_h^{EL}) - \frac{GEN_{h,t=1}}{\sigma_h^{OCGT}} \ : \phi_{h,t=1}^{LVLH}, \forall h \tag{21}$$

$$LVL_{h,t} = LVL_{h,t-1} + (CH_{h,t} \cdot \sigma_h^{EL}) - \frac{GEN_{h,t}}{\sigma_h^{OCGT}} \ : \phi_{h,t}^{LVLH}, \forall h, \forall t>1, \tag{22}$$

$$PF_{l,t} = sus_l \cdot \left(\theta_{s(l)} - \theta_{r(l)}\right) \ : \phi_{l,t}^{L}, \forall l,t \tag{23}$$

$$-\left(\overline{cap}_l + CAP_l\right) \leq PF_{l,t} \leq \overline{cap}_l + CAP_l \ : \ \underline{\mu}_{l,t}^{L}, \overline{\mu}_{l,t}^{L}, \forall l,t \tag{24}$$

$$-\pi \leq \theta_{n,t} \leq \pi \ : \ \underline{\mu}_{n,t}^{N}, \overline{\mu}_{n,t}^{N}, \forall n,t \tag{25}$$

$$\theta_{n,t} = 0 : \epsilon^{ref}, \ n \in ref(n) \tag{26}$$

$$LS_{n,t}^{L1} \leq dem_{n,t} \cdot f^{L1} \ : \mu_{n,t}^{LS1}, \forall n,t \tag{27}$$

$$LS_{n,t}^{L2} \leq dem_{n,t} \cdot f^{L2} \ : \mu_{n,t}^{LS2}, \forall n,t \tag{28}$$

$$LS_{n,t}^{L3} \leq dem_{n,t} \cdot f^{L3} \ : \mu_{n,t}^{LS3}, \forall n,t \tag{29}$$

The objective function in Eq. (1) minimizes the sum of the annualized renewable capacity and transmission investment costs, the annualized costs of the battery storage systems, which includes inverter and storage cost components, the annualized costs of the hydrogen storage systems, which include the electrolyzer, the $H_2$ storage tank and the $H_2$ fired open cycle gas turbines (OCGTs) power plant, as well as the variable generation costs and the load shedding costs. The minimization problem is subject to several constraints such as the energy balance in Eq. (2), the technical generation constraints of renewable generation units, considering the capacity factors for solar PV and wind ($\widetilde{cf}_{pv,t}^{M}, \widetilde{cf}_{wind,t}^{M}$) in Eqs. (3-4), the conventional generation in Eq. (5), and hydropower generation in Eqs. (6-7), the storage constraints for pumping storage plants in Eqs. (8-12), battery storage systems in Eqs. (13-17) and hydrogen storage in Eqs. (18-22). The model does not consider an underlying hydrogen transmission network. Therefore, hydrogen storage tanks located at a specific node can only be charged or discharged by collocated electrolyzer and hydrogen-fired OCGT power plants that consume electricity or hydrogen, respectively.

We consider DC power flow constraints in Eqs. (23-26), and three load shedding constraints in Eqs. (27-29). Note that these three load shedding constraints establish a stepwise load shedding curve, where the factors $f^{L1}, f^{L2}, f^{L3}$ (with $f^{L1} < f^{L2} < f^{L3}$) represent specific percentages of demand at a given node and time. The load shedding variables $LS_{n,t}^{L1}, LS_{n,t}^{L2}, LS_{n,t}^{L3}$ are multiplied with different cost levels in the objective function. This implementation creates a stepwise load shedding cost function, enabling the model to utilize varying degrees of load flexibility at different cost levels.

## 3.2 Worst case realizations and corrective actions (Subproblem)

Our approach to modeling uncertainty as the realization of low solar PV and wind availability involves defining cardinality-constrained uncertainty sets for these renewable generation technologies. To avoid confusion between general descriptive language and model-specific terminology in the reminder, we introduce the following distinction: the term extreme weather event is used generically to describe scarcity periods of significantly reduced renewable generation, whether due to low wind, low solar, or both. In contrast, a worst-case renewable scarcity event refers to a specific uncertainty realization identified within our adaptive robust optimization framework, representing the most adverse combination of low renewable availability for system planning.

The cardinality-constrained uncertainty set $\Phi^U$ controls the number of uncertain parameters that can deviate from their nominal values. This is controlled by binary variables $Z$ in combination with the uncertainty budget $\Gamma$, and can be described formally as follows:

$$\Phi^U = \left\{ \begin{array}{lr} \widetilde{CF}_{r,g,t} = \overline{cf}_{r,g,t} - \sum_{p \in P(t)} Z_{r,g,p} \cdot \widehat{cf}_{r,g,t} & \forall r, \forall g, \forall t, \forall p \quad (30) \\ Z_{r,g,p} \in \{0,1\} & \forall r, \forall g, \forall p \quad (31) \\ \sum_{g \in G} Z_{r,g,p} \leq \Gamma_{r,p} & \forall r, \forall p \quad (32) \end{array} \right\}$$

Equation (30) describes the uncertain renewable capacity factor variable $\widetilde{CF}_{r,g,t}$ for a specific renewable technology $r$ (solar PV or wind), located in weather region $g$ and time step $t$ depending on the reference value $\overline{cf}_{r,g,t}$ (expected historical average) and the deviation values $\widehat{cf}_{r,g,t}$. The realization of the deviation depends on the value of the binary variable $Z_{r,g,p}$, defined by eq. (31). Note here, that the binary variable is defined over a certain period $p$. The period $p$ is defining a time horizon, spanning over several timesteps, such as a day, week, or month including several hours. The Subproblem then calculates both the most unfortunate uncertainty realizations with respect to the variables in the set $\Phi^U$ and the corrective dispatch actions of variables in set $\Phi^S = \{\mu_{c,t}^{Conv}; \mu_{pv,t}^{R}; \mu_{wind,t}^{R}; \mu_{ror,t}^{ROR}; \mu_{rsv,t}^{RSV}; \mu_{PSP,t}^{DIS}; \mu_{psp,t}^{CH}; \underline{\mu}_{b,t}^{CAPB}; \mu_{h,t}^{EL}; \mu_{h,t}^{OCGT}; \underline{\mu}_{h,t}^{CAPH}; \mu_{n,t}^{LS1}; \mu_{n,t}^{LS2}; \mu_{n,t}^{LS3}; \overline{\mu}_{l,t}^{L}; \underline{\mu}_{l,t}^{L}; \underline{\mu}_{n,t}^{N}; \phi_{pv,t}^{aux}; \phi_{wind,t}^{aux}; \lambda_{n,t}; \phi_{l,t}^{L}; \phi_{PSP,t}^{LVL}; \phi_{b,t}^{LVLB}; \phi_{h,t}^{LVLH}; \epsilon^{ref}\}$, maximizing the total system cost. Set $\Phi^S$ is also referred to as feasibility set. A mathematical description of how the Subproblem in general is derived can be found in Appendix A.

Note that the model endogenously and solely computes the worst-case scarcity event realizations based on the impact on the total system cost. On this basis, the model can consider either scarcity events of a single renewable technology in a specific region or the simultaneous scarcity of both. If both wind and solar availability are simultaneously reduced, we refer to this as a worst-case "Dunkelflaute" event.

The resulting Subproblem takes the following form:

$$\underset{\Phi^U \Phi^S}{Max}: \tag{33}$$

$$\begin{aligned} Objective^{SUB} = & \sum_{n,t} \lambda_{n,t} \cdot dem_{n,t} + \sum_{c,t} -\mu_{c,t}^{Conv} \cdot \overline{cap}_c + \sum_{pv,t} -\mu_{pv,t}^{R} \cdot (\overline{\text{sub}}_{pv}^{REN} \cdot \overline{cf}_{pv,t}) + \phi_{pv,t}^{aux} \\ & \cdot (\overline{\text{sub}}_{pv}^{REN} \cdot \widehat{cf}_{pv,t}) + \sum_{wind,t} -\mu_{wind,t}^{R} \cdot (\overline{\text{sub}}_{wind}^{REN} \cdot \overline{cf}_{wind,t}) + \phi_{wind,t}^{aux} \cdot (\overline{\text{sub}}_{wind}^{REN} \\ & \cdot \widehat{cf}_{wind,t}) + \sum_{ror,t} -\mu_{ror,t}^{ROR} \cdot \overline{cap}_{ror} + \sum_{rsv,t} -\mu_{rsv,t}^{RSV} \cdot (\overline{cap}_{rsv} \cdot af_{rsv,t}) + \sum_{psp,t} -\mu_{psp,t}^{DIS} \\ & \cdot \overline{cap}_{psp} - \mu_{psp,t}^{CH} \cdot \overline{cap}_{psp} - \mu_{PSP,t}^{CAP} \cdot (\overline{cap}_{psp} \\ & \cdot csf_{psp}) + \sum_{psp,t=1} \phi_{PSP,t=1}^{LVL} \\ & \cdot \frac{\overline{cap}_{psp} \cdot csf_{psp}}{2} + \sum_{b,t} -\mu_{b,t}^{DB} \cdot \overline{\text{sub}}_{b}^{INV} - \mu_{b,t}^{CB} \\ & \cdot \overline{\text{sub}}_{b}^{INV} - \mu_{b,t}^{CAPB} \cdot \overline{\text{sub}}_{b}^{STOR} + \sum_{h,t} -\mu_{h,t}^{OCGT} \cdot \overline{\text{sub}}_{h}^{OCGT} - \mu_{h,t}^{EL} \cdot \overline{\text{sub}}_{h}^{EL} - \mu_{h,t}^{CAPH} \\ & \cdot \overline{\text{sub}}_{h}^{STOR} + \sum_{n,t} -\mu_{n,t}^{LS1} \cdot (dem_{n,t} \cdot f^{L1}) + \sum_{n,t} -\mu_{n,t}^{LS2} \cdot (dem_{n,t} \\ & \cdot f^{L2}) + \sum_{n,t} -\mu_{n,t}^{LS3} \cdot (dem_{n,t} \cdot f^{L3}) + \sum_{l,t} -\underline{\mu}_{l,t}^{L} \cdot \overline{\text{sub}}_{l}^{Line} + \sum_{l,t} -\overline{\mu}_{l,t}^{L} \cdot \overline{\text{sub}}_{l}^{Line} \end{aligned}$$

$$\widetilde{CF}_{pv,g,t} = \overline{cf}_{pv,g,t} - \sum_{p \in P(t)} Z_{pv,g,p} \cdot \widehat{cf}_{pv,g,t} \quad \forall pv(r), \forall g, \forall t, \forall p \tag{34}$$

$$\widetilde{CF}_{wind,g,t} = \overline{cf}_{wind,g,t} - \sum_{p \in P(t)} Z_{wind,g,p} \cdot \widehat{cf}_{wind,g,t} \quad \forall wind(r), \forall g, \forall t, \forall p \tag{35}$$

$$\sum_{g \in G} Z_{pv,g,p} \leq \Gamma_{pv,p} \quad \forall pv(r), \forall p \tag{36}$$

$$\sum_{g \in G} Z_{wind,g,p} \leq \Gamma_{wind,p} \quad \forall wind(r), \forall p \tag{37}$$

$$\lambda_{n,t} - \mu_{c,t}^{Conv} \leq 0, \ \forall n, \forall c, \forall t \tag{38}$$

$$\lambda_{n,t} - \mu_{pv,t}^{R} \leq 0, \ \forall n, \forall pv(r), \forall t \tag{39}$$

$$\lambda_{n,t} - \mu_{wind,t}^{R} \leq 0, \ \forall n, \forall wind(r), \forall t \tag{40}$$

$$\lambda_{n,t} - \mu_{ror,t}^{ROR} \leq 0, \ \forall n, \forall ror, \forall t \tag{41}$$

$$\lambda_{n,t} - \mu_{rsv,t}^{RSV} \leq 0, \ \forall n, \forall rsv, \forall t \tag{42}$$

$$\lambda_{n,t} - \phi_{PSP,t}^{LVL} - \mu_{PSP,t}^{DIS} \leq 0, \ \forall n, \forall psp, \forall t \tag{43}$$

$$-\lambda_{n,t} + \phi_{PSP,t}^{LVL} \cdot \sigma_{psp} - \mu_{psp,t}^{CH} \leq 0, \ \forall n, \forall psp, \forall t \tag{44}$$

$$\phi_{PSP,t}^{LVL} - \phi_{PSP,t\text{-}1}^{LVL} - \mu_{PSP,t}^{CAP} + \underline{\mu}_{PSP,t}^{CAP} = 0, \quad \forall psp, \forall t \tag{45}$$

$$\phi_{PSP,t=1}^{LVL} - \mu_{PSP,t=1}^{CAP} + \underline{\mu}_{PSP,t=1}^{CAP} = 0, \quad \forall psp \tag{46}$$

$$\lambda_{n,t} - \phi_{b,t}^{LVLB} - \mu_{b,t}^{DB} \leq 0, \ \forall n, \forall b, \forall t \tag{47}$$

$$-\lambda_{n,t} + \phi_{b,t}^{LVLB} \cdot \sigma_b - \mu_{b,t}^{CB} \leq 0, \ \forall n, \forall b, \forall t \tag{48}$$

$$-\phi_{b,t}^{LVLB} + \phi_{b,t+1}^{LVLB} - \mu_{b,t}^{CAPB} + \underline{\mu}_{b,t}^{CAPB} = 0, \ \forall b, \forall t \in Tt_{last}\} \tag{49}$$

$$-\phi_{b,t}^{LVLB} - \mu_{b,t}^{CAPB} + \underline{\mu}_{b,t}^{CAPB} = 0, \ \forall b, \forall t_{last} \tag{50}$$

$$\lambda_{n,t} - \frac{\phi_{h,t}^{LVLH}}{\sigma_h^{OCGT}} - \mu_{h,t}^{OCGT} \leq 0, \ \forall n, \forall h, \forall t \tag{51}$$

$$-\lambda_{n,t} + \phi_{h,t}^{LVLH} \cdot \sigma_h^{EL} - \mu_{h,t}^{EL} \leq 0, \ \forall n, \forall h, \forall t \tag{52}$$

$$-\phi_{h,t}^{LVLH} + \phi_{h,t+1}^{LVLH} - \mu_{h,t}^{CAPH} + \underline{\mu}_{h,t}^{CAPH} = 0, \ \forall h, \forall t \in Tt_{last}\} \tag{53}$$

$$-\phi_{h,t}^{LVLH} - \mu_{h,t}^{CAPH} + \underline{\mu}_{h,t}^{CAPH} = 0, \ \forall h, \forall t_{last} \tag{54}$$

$$\lambda_{n,t} - \mu_{n,t}^{LS1} \leq sc_n^{LS1}, \ \forall n, \forall t \tag{55}$$

$$\lambda_{n,t} - \mu_{n,t}^{LS2} \leq sc_n^{LS2}, \ \forall n, \forall t \tag{56}$$

$$\lambda_{n,t} - \mu_{n,t}^{LS3} \leq sc_n^{LS3}, \ \forall n, \forall t \tag{57}$$

$$-\lambda_{s(l)} + \lambda_{r(l)} - \overline{\mu}_{l,t}^{L} + \underline{\mu}_{l,t}^{L} + \phi_{l,t}^{L} = 0, \ \forall t, \forall l \in AC(l) \tag{58}$$

$$-\lambda_{s(l)} + \lambda_{r(l)} - \overline{\mu}_{l,t}^{L} + \underline{\mu}_{l,t}^{L} = 0, \ \forall t, \forall l \in DC(l) \tag{59}$$

$$\sum_{s(l)} (sus_l \cdot \phi_{l,t}^{L}) - \sum_{r(l)} (sus_l \cdot \phi_{l,t}^{L}) - \underline{\mu}_{n,t}^{N} + \overline{\mu}_{n,t}^{N} = 0, \ \forall n, \forall t \tag{60}$$

$$\sum_{s(l)} (sus_l \cdot \phi_{l,t}^{L}) - \sum_{r(l)} (sus_l \cdot \phi_{l,t}^{L}) - \underline{\mu}_{n,t}^{N} + \overline{\mu}_{n,t}^{N} + \epsilon^{ref} = 0, \ \forall n, \forall t \tag{61}$$

$$(-M) \cdot Z_{pv,g,p} \leq \phi_{pv,t}^{aux} \leq M \cdot Z_{pv,g,p}, \ \forall pv(r), \forall g, \forall p, \forall t \tag{62}$$

$$(-M) \cdot \left(1 - Z_{pv,g,p}\right) \leq \mu_{pv,t}^{R} - \phi_{pv,t}^{aux} \leq M \cdot \left(1 - Z_{pv,g,p}\right), \quad \forall pv(r), \forall g, \forall p, \forall t \tag{63}$$

$$(-M) \cdot Z_{wind,g,p} \leq \phi_{wind,t}^{aux} \leq M \cdot Z_{wind,g,p}, \quad \forall wind(r), \forall g, \forall p, \forall t \tag{64}$$

$$(-M) \cdot \left(1 - Z_{wind,g,p}\right) \leq \mu_{wind,t}^{R} - \phi_{wind,t}^{aux} \leq M \cdot \left(1 - Z_{wind,g,p}\right), \quad \forall wind(r), \forall g, \forall p, \forall t \tag{65}$$

**Positive variables:** $\mu_{c,t}^{Conv}; \mu_{pv,t}^{R}; \mu_{wind,t}^{R}; \mu_{ror,t}^{ROR}; \mu_{rsv,t}^{RSV}; \mu_{PSP,t}^{DIS}; \mu_{psp,t}^{CH}; \underline{\mu}_{b,t}^{CAPB}; \mu_{h,t}^{EL}; \mu_{h,t}^{OCGT}; \underline{\mu}_{h,t}^{CAPH};$ (66)

$\mu_{n,t}^{LS1}; \mu_{n,t}^{LS2}; \mu_{n,t}^{LS3}; \overline{\mu}_{l,t}^{L}; \underline{\mu}_{l,t}^{L}; \overline{\mu}_{n,t}^{N}; \underline{\mu}_{n,t}^{N}; \phi_{pv,t}^{aux}; \phi_{wind,t}^{aux}$

**Free variables:** $\lambda_{n,t}; \phi_{l,t}^{L}; \phi_{PSP,t}^{LVL}; \phi_{b,t}^{LVLB}; \phi_{h,t}^{LVLH}; \epsilon^{ref}$ (67)

Note that within the formulation of the Master Problem, the renewable availability constraints (3) and (4) are determined as variables by the Subproblem which maximizes the objective function by identifying the most unfortunate realization of $\widetilde{CF}_{pv,g,t}$ and $\widetilde{CF}_{wind,g,t}$, defined by Eqs. (34)-(37). The dual generation constraints of the conventional generators, the renewable generators, and the hydro generators are described by Eqs. (38)-(42). The dual storage reformulation for the pumped hydro storage is defined by Eqs. (43-46), followed by the dual battery storage formulation in Eqs. (47)-(50), and by the dual hydrogen storage formulation from Eqs. (51)-(54). The dual load shedding formulation is described by Eqs. (55)-(57), followed by the dual formulation of the transmission line flows in Eqs. (58)-(61). We use the big M constraints in Eqs. (62)-(65) and introduce the auxiliary variables $\phi_{pv,t}^{aux}$ and $\phi_{wind,t}^{aux}$ to replace the non-linear term in the objective function $\mu_{pv,t}^{R} \cdot \widetilde{CF}_{pv,g,t}$and $\mu_{wind,t}^{R} \cdot \widetilde{CF}_{wind,g,t}$ with an equivalent linear formulation. The dual variable types are declared in Eqs. (66) and (67).

## 3.3 Solution Strategy

The problem is solved with a column-constraint generation algorithm that iterates over the Master Problem and Subproblems until a predefined convergence tolerance $\epsilon$ between the two objective functions is archived. During this process, the Master Problem transfers specific information on expansion of infrastructure assets such as generation, storage, and transmission in set $\Phi^{M \to S}$ to the Subproblem, which sends back information about worst case realizations and corrective actions in set $\Phi^{S \to M}$ to the Master Problem. This process is visualized in Figure 1 below.

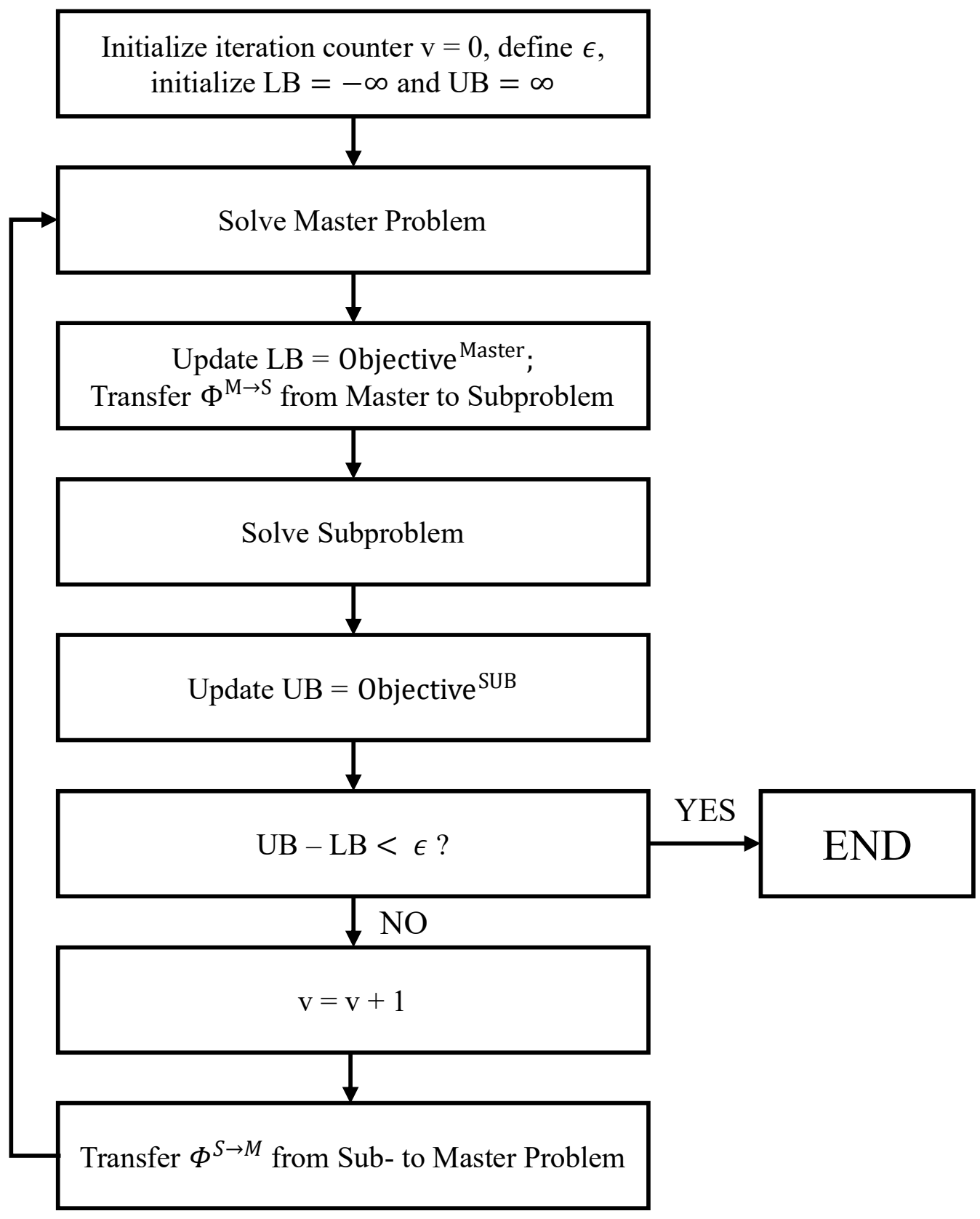

**Figure 1:** Solution strategy – column and constrain generation algorithm based on Baringo and Rahimiyan, (2020)

The algorithm starts with defining the iteration counter, the tolerance level $\epsilon$ and the initialization of the lower and upper bounds LB and UB. Next, the Master Problem is solved. Then, the LB parameter is updated and the generation and transmission capacity expansion decisions are transferred to the Subproblem by storing the information in the parameter set $\Phi^{M \to S}$={$\overline{\text{sub}}_{pv}^{REN}$ ; $\overline{\text{sub}}_{wind}^{REN}$; $\overline{\text{sub}}_{b}^{INV}$ ; $\overline{\text{sub}}_{b}^{STOR}$ ; $\overline{\text{sub}}_{h}^{OCGT}$ ; $\overline{\text{sub}}_{h}^{EL}$ ; $\overline{\text{sub}}_{h}^{STOR}$ ; $\overline{\text{sub}}_{l}^{Line}$ }. On this basis the corresponding Subproblem is solved followed by an update of the UB parameter. As a next step, the convergence criterion is checked. If $UB - LB < \epsilon$ the algorithm stops. The optimal expansion decisions are within set $\Phi^{M \to S}$. If the convergence criterion is not met, the iteration counter is updated. The information of the renewable capacity factors worst case realizations which are stored in the Subproblem in parameters $\widetilde{CF}_{pv,g,t}$ and $\widetilde{CF}_{wind,g,t}$ , are then transferred to the Master Problem, described by set $\Phi^{S \to M}$={$\widetilde{CF}_{pv,g,t} \to \widetilde{cf}_{pv,t}^{M}$; $\widetilde{CF}_{wind,g,t} \to \widetilde{cf}_{wind,t}^{M}$}. The Master Problem is re-solved incorporating both (if available) previously identified and newly generated worst-case realizations. With each iteration, the Master Problem grows (augmented by additional cuts), progressively accounting for adverse uncertainty realizations. This iterative process continues until the convergence criterion is satisfied, at which point the optimal robust expansion decisions are obtained.

# 4 APPLICATION – CASE STUDY

For our study we develop a generation and transmission expansion planning model for a 100 % renewable electricity system for 24 European countries with the planning horizon of 2050. To represent electricity flows, we use PyPSA-Eur to cluster the ENTSO-E transmission system data and to generate a representative high-voltage electricity network. This electricity network spans all considered countries and consists of 50 nodes and 97 transmission lines and links as depicted in Figure 2. The model differentiates expandable and non-expandable generation technologies. Expandable generation technologies allow endogenous investment in new capacity comprising solar PV, onshore wind, offshore wind (where coastal connections exist), hydrogen storage, electrolyzers, hydrogen-fired open cycle gas turbines (OCGT), battery storage, and inverters. Also, the transmission line capacity can be expanded. However, expansion is limited to an additional 100 % of 2020 levels. While net-zero pathways may benefit from more extensive grid expansion, this assumption takes into account that transmission projects at the European level face substantial social, political, and regulatory constraints. Non-endogenous capacities comprise hydro and nuclear power. We fix hydro power, which includes run-of-river (RoR), reservoir and pumping storage power plants, at their current levels since we assume that expansion potential is negligible. For nuclear power capacity in 2050, we consider existing power plants that will still be in operation in 2050, given a lifetime of 60 years, as well as newly planned nuclear power projects based on data from the World Nuclear Association. Details on the specific technology data and costs are presented in Appendix B.

The electricity demand for each node in 2050 was estimated using the open-source GlobalEnergyGIS package based on Mattsson et al. (2021). First, we projected annual electricity consumption for each region in 2050 based on 2016 demand data taken from the International Energy Agency, and regional growth in the Shared Socioeconomic Pathway 2 scenario from Riahi et al. (2017). Then, we estimated the hourly demand profile using a machine learning model that applies historical demand data from 44 countries to a gradient boosting regression approach based on Friedman (2001). This model incorporates calendar effects (e.g. hour of day, weekday/weekend), temperature (e.g. hourly temperature in key population areas), and economic indicators (e.g. local gross domestic product per capita). Finally, the hourly demand series is scaled to match the projected annual demand for each region in 2050.

To incorporate weather data, we analyzed 40 historical weather years based on data from the Renewable.ninja website (data are based on the works of Pfenninger and Staffell (2016), and Staffell and Pfenninger (2016)). For each node in the system, we include individual historical average capacity factor time series for solar PV, and onshore and offshore wind technologies. To improve the tractability of the model, we apply a time series reduction method (moving average) to the original time series of hourly resolved demand and renewable capacity factors, reducing them to a time series with four-hour timesteps. In the following section, we describe the parameterization of the weather-related uncertainty and present the scenarios of the case study that determine the geographical coverage of the wind and solar scarcity events. To support transparency and reproducibility, all data, code, and results are publicly available on GitHub[3], allowing readers to verify, adapt, or extend our work.

[3] https://github.com/bernemax/ARO_Dunkelflaute_Europe

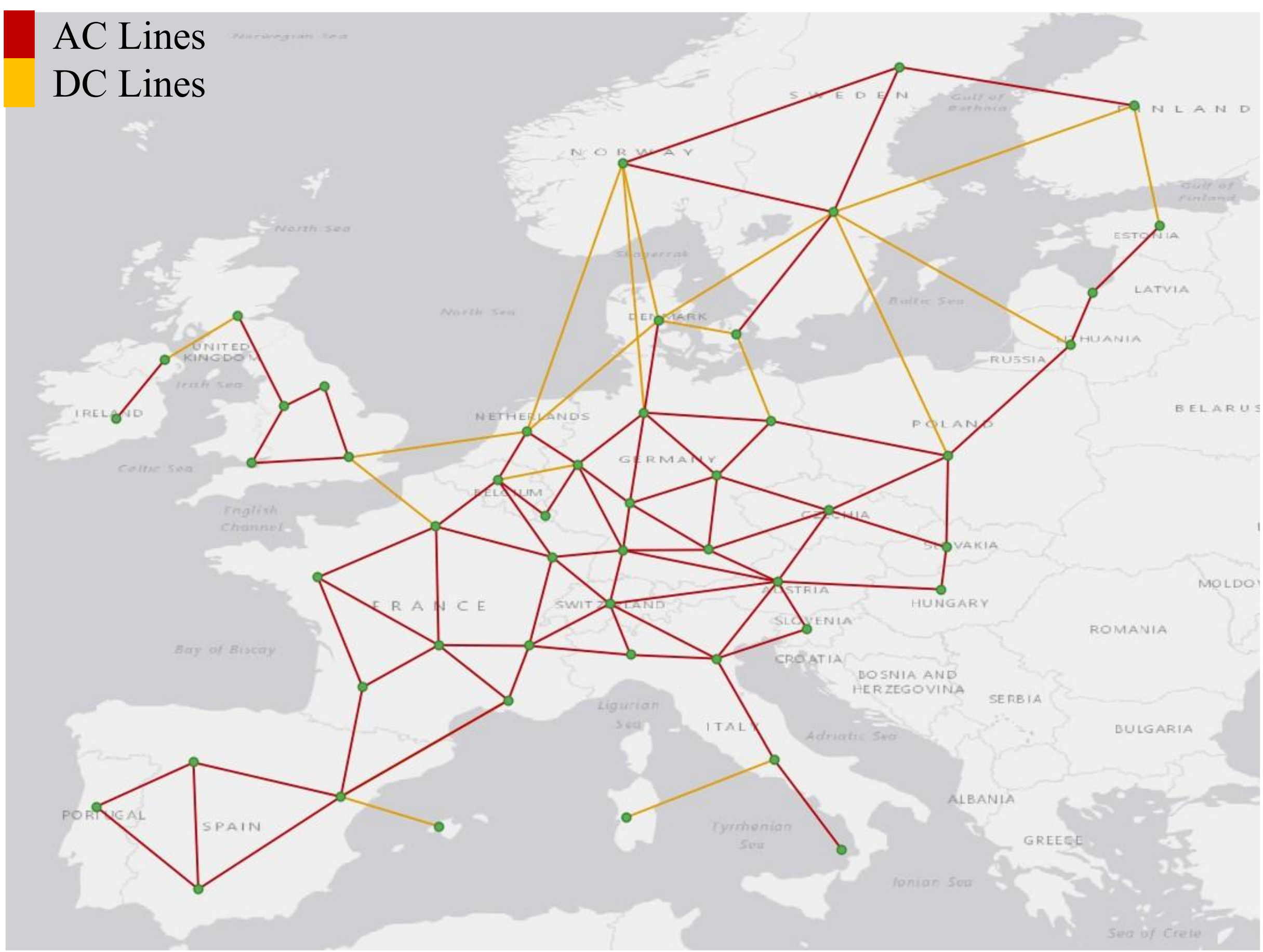

**Figure 2:** Map of clustered EU high-voltage electricity system – own illustration

## 4.1 Uncertainty Sets

To assess which combination of generation, storage, and transmission technologies ensures the robustness of a future European electricity system against worst-case realizations of low availability events, we stress-tested the expansion model using synthetic extreme weather scenarios. Our methodology for modeling these worst-case scarcity events consisted of two main steps. In the first step, we defined the geographical scope, introducing weather regions within Europe. In these regions, solar and wind availability can be modeled either as typical availability conditions or as scarcity periods consisting of extreme low availability scenarios. In the second step, we defined the uncertainty realization space by specifying the number of affected regions, using a cardinality-constrained uncertainty set. Note, that the cardinality-constrained uncertainty set is not intended to reproduce physically continuous or fully correlated meteorological events. Instead, it serves as a planning-oriented abstraction that captures the system's exposure to simultaneous renewable scarcity events across an increasing number of regions. By constraining uncertainty through the number of affected regions rather than the exact intensity or spatio-temporal structure of weather events, the model focuses on identifying structural vulnerabilities and capacity adaptation mechanisms under progressively adverse conditions. The use of historical lower-bound availability profiles represents an extreme, ensuring that the system is robust against severe scarcity events without relying on specific historical realizations. This abstraction deliberately sacrifices meteorological detail in favor of computational tractability and transparency, enabling the identification of worst-case regions and technologies that drive system adaptation costs. In the following, we outline our methodology, first addressing the geographic scope of the extreme scarcity events, and then describing the cardinality-constrained uncertainty set.

### 4.1.1 Geographical Scope

In order to incorporate the effects of weather events with low solar and wind availability, our approach relied on splitting the modeled European system into six weather regions. Therefore, we assigned 24 European countries to these regions in accordance with the study by Otero et al., (2022), as shown in Figure 3. Those authors analyzed the correlation between the monthly frequencies of energy compound events (periods with low wind and solar irradiance) among different European countries, considering the spatial relationship between critical energy events across Europe. Within these regions we assume that the weather conditions, which translate back to the nodal renewable capacity factors, are the same: normal availability or low solar availability or low wind availability, or both low solar availability and low wind availability.

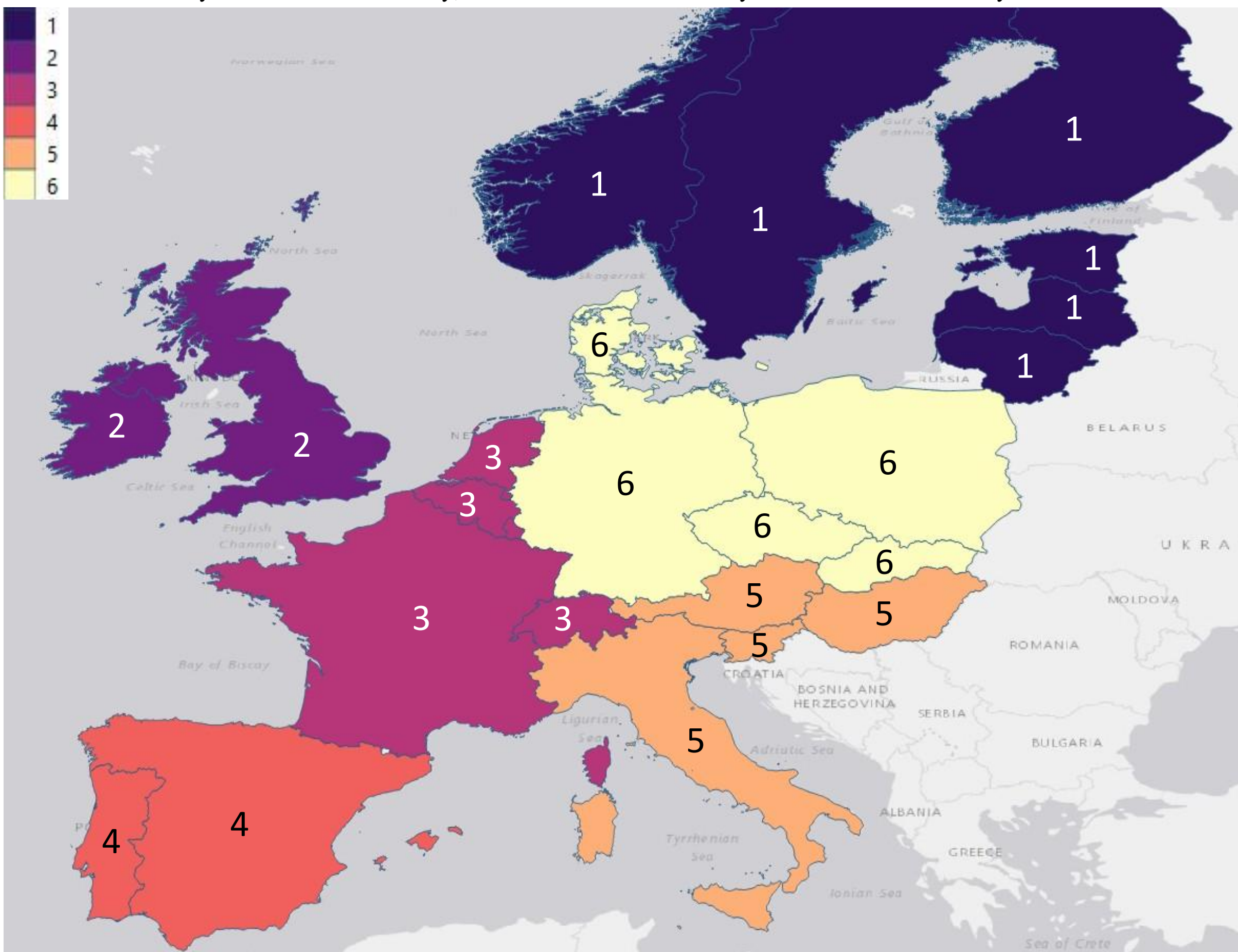

**Figure 3:** Map of weather regions, Region 1: NO, SE, FI, LV, LT, EE, Region 2: GB, IE, Region 3: NL, BE, FR, CH, LU, Region 4: ES, PT, Region 5: IT, AT, SI, HU, Region 6: DE, DK, CZ, PL, SK– own illustration.

#### 4.1.2 Renewable Uncertainty - Capacity Factors Lower Bounds

In our case study, we set the renewable scarcity event duration to be one week (7 days), which corresponds also to a typical time span of a worst case Dunkelflaute event (Raynaud et al., 2018; Li et al., 2021b; Otero et al., 2022b). The uncertainty realization of such an extreme event is controlled by the uncertainty budget $\Gamma_{r,p}$ as defined in eq. (32). The budget controls the number of deviations in solar or wind availability from their reference values across the defined weather regions and time periods. Here $\Gamma_{r,p}$=0 implies that no uncertainty occurs for the specific generation technology, meaning that the renewable capacity factor time series remains at its expected reference values for all regions. For all discrete values $\Gamma_{r,g} \geq 0$, a low renewable availability (scarcity) period can occur in a specific weather region as the deviation value $\widehat{cf}_{r,g,t}$ is subtracted from the reference value $\overline{cf}_{r,g,t}$. To determine $\widehat{cf}_{r,g,t}$, we generated a synthetic low-capacity factor time series which serves as a so-called lower bound. Using 40 historical weather years, we first compute weekly average capacity factors for wind and solar generation. For each calendar week, we then identify the historical week with the lowest average capacity factor (separately for each technology). The original 52 selected low capacity factor weeks are concatenated to form a full synthetic "worst-case" year representing a lower-bound availability profile. The deviation value $\widehat{cf}_{r,g,t}$ then corresponds to the difference between the reference value $\widetilde{cf}_{r,g,t}$ and the lower bound of the capacity factor. An exemplary visualization of the lower bound and the reference values for solar PV in Austria is presented in Appendix C.

## 4.2 Scenario Definition

Since we examine the effect of different geographical coverages of low availability events, we investigate six different ARO scenarios, by gradually increasing the uncertainty budgets $\Gamma_{PV}$ and $\Gamma_{Wind}$ from 1 to 6. In scenario ARON, exactly N regions experience a low-availability event for wind, and N regions experience such an event for solar PV. Accordingly, ARO1 through ARO6 represent cases with one to six affected regions, respectively. More formally, the integer number of the respective uncertainty budget represents the number of regions that can be affected by low solar PV or wind availability periods. Thus, the model endogenously decides in which region the availability is reduced. For example, in scenario ARO1, one region can be affected by low availability for both solar PV and wind, in the ARO2 scenario, the model can set the availability of solar PV and wind to the lower bound in two regions and so on, up to ARO6, where all six regions can simultaneously experience low availability—effectively representing a pan-European Dunkelflaute. We contrast these scenarios to a Base scenario, in which no low availability event realizes. We restrict the model to simulate weekly low solar and wind generation periods in January, as climatological studies indicate that the most severe Dunkelflaute events predominantly occur in mid-winter, when low renewable availability coincides with high electricity demand, making January a conservative and system-stressing reference period (H. C. Bloomfield et al., 2020; Li et al., 2021a; Grochowicz et al., 2024).

# 5 MODEL RESULTS

In the following, we outline the application of the ARO framework and the results of our case study. Section 5.1 presents the iterative modeling process, with a focus on the convergence behavior of the min-max problem and the extreme scarcity event realizations. Section 5.2 presents the identified worst-case weather events and analyzes their impact on investment and generation costs. Section 5.3 examines the resulting robust, technology-specific generation mix, while Section 5.4 discusses the corresponding storage capacities for each scenario, both at the European level and disaggregated by region.

## 5.1 Iterative ARO Process – Incorporation of Uncertainty

The model endogenously stresses the specified weather regions with low solar PV or wind availability events. Therefore, it iterates over the possible realization space which includes four weeks of January and the six geographical regions[4]. For each scenario (ARO1-ARO6), the Table 1 below indicates the iteration (it1-it14), region (1-6), and time period (four weeks in January) for which the renewable scarcity event realization can occur. The colored cells highlight the times and regions at which the capacity factor time series is reduced to their historical lower bounds, where yellow is allocated for solar PV, blue for onshore and offshore wind, and black for both technologies simultaneously (Dunkelflaute). For example, in scenario ARO1, the model explores scarcity event realization in 12 iterations, starting with low solar PV availability in week 4 of January in Region 4 and reduced wind availability in Region 5. Subsequent iterations test other combinations, such as a Dunkelflaute (simultaneous low solar PV and wind availability) in week 2 in Region 3. The red-framed cells in the table indicate the iteration at which the model converges, presenting our final worst-case solution. Note that the capacity layout for the resulting system is robust towards all combinations possible, not only against the ones tested. Robustness in the ARO framework is enforced through the definition of the uncertainty set rather than by enumerating individual scenarios. This means that the optimization problem is formulated over all admissible realizations of renewable availability, not only over a finite set of explicitly tested scenarios. From an optimization perspective, only uncertainty realizations that are binding for the feasibility or are cost optimal influence the solution. Realizations that induce lower system stress than the worst case do not affect the optimal investment decision and are therefore irrelevant for determining the robust capacity layout. Moreover, the solution is not tailored to specific scenarios, but to the worst-case realization within the uncertainty set. The model therefore converges to a solution that is driven by the most critical realizations, which implicitly cover the full uncertainty set. Occasionally, the same extreme event reappears across iterations - e.g., in ARO1, the Dunkelflaute in Region 3 during week 2 is identified in both iteration 2 and the final iteration 12. This occurs because the model retains all previous iterations when looking for the worst-case event and adopts the generation and transmission expansion decisions accordingly. Consequently, the memory of the model grows with each iteration. Thus, identifying a possible worst-case event in iteration 3 is easier than in iteration 12. In some cases, after evaluating the previous possibilities, the model reconfirms the same event again to be a possible worst case. This means that the generation and transmission expansion adoptions made lead the model to find the event realization as only possibility, where the primal objective equals the dual objective. All other uncertainty realizations are automatically excluded, and the model converges.

As the uncertainty budget increases from the ARO1 to the ARO6 scenario, more regions are affected either simultaneously or even at different times. Overall, the model finds simultaneous extreme events to be harder to mitigate. For instance, in scenario ARO4, iteration 1 features solar and wind scarcity events both week 3 and week 4. In subsequent iterations, only simultaneous scarcity periods occur, with varying low availability events between the affected regions. Furthermore, one can see that the number of cells that show both technologies affected (Dunkelflaute) increases with higher uncertainty budget.

[4] The model terminates when the gap between the primal and the dual objective value is lower than a predefined tolerance. For our case the threshold is less than 0.000001% of the objective function value. This process is illustrated in more detail in Appendix .

**Table 1: Iterative uncertainty selection process of the model for each scenario. The colors represent the individual technology where the capacity factor is reduced due to a low availability event, namely yellow for solar PV, blue for wind and black for both technologies simultaneously.**

| Iteration | Week January | ARO1 | | | | | | ARO2 | | | | | | ARO3 | | | | | | ARO4 | | | | | | ARO5 | | | | | | ARO6 | | | | | |
|---|---|---|---|---|---|---|---|---|---|---|---|---|---|---|---|---|---|---|---|---|---|---|---|---|---|---|---|---|---|---|---|---|---|---|---|---|---|
| | | Regions | | | | | | Regions | | | | | | Regions | | | | | | Regions | | | | | | Regions | | | | | | Regions | | | | | |
| | | 1 | 2 | 3 | 4 | 5 | 6 | 1 | 2 | 3 | 4 | 5 | 6 | 1 | 2 | 3 | 4 | 5 | 6 | 1 | 2 | 3 | 4 | 5 | 6 | 1 | 2 | 3 | 4 | 5 | 6 | 1 | 2 | 3 | 4 | 5 | 6 |
| it1 | 1 | | | | | | | | | | | | | | | | | | | | | | | | | | | | | | | | | | | | |
| | 2 | | | | | | | | | | | | | | | | | | | | | | | | | | | | | | | | | | | | |
| | 3 | | | | | | | | | | | | | | | | | | | | | | | | | | | | | | | | | | | | |
| | 4 | | | | | | | | | | | | | | | | | | | | | | | | | | | | | | | | | | | | |
| it2 | 1 | | | | | | | | | | | | | | | | | | | | | | | | | | | | | | | | | | | | |
| | 2 | | | | | | | | | | | | | | | | | | | | | | | | | | | | | | | | | | | | |
| | 3 | | | | | | | | | | | | | | | | | | | | | | | | | | | | | | | | | | | | |
| | 4 | | | | | | | | | | | | | | | | | | | | | | | | | | | | | | | | | | | | |
| it3 | 1 | | | | | | | | | | | | | | | | | | | | | | | | | | | | | | | | | | | | |
| | 2 | | | | | | | | | | | | | | | | | | | | | | | | | | | | | | | | | | | | |
| | 3 | | | | | | | | | | | | | | | | | | | | | | | | | | | | | | | | | | | | |
| | 4 | | | | | | | | | | | | | | | | | | | | | | | | | | | | | | | | | | | | |
| it4 | 1 | | | | | | | | | | | | | | | | | | | | | | | | | | | | | | | | | | | | |
| | 2 | | | | | | | | | | | | | | | | | | | | | | | | | | | | | | | | | | | | |
| | 3 | | | | | | | | | | | | | | | | | | | | | | | | | | | | | | | | | | | | |
| | 4 | | | | | | | | | | | | | | | | | | | | | | | | | | | | | | | | | | | | |
| it5 | 1 | | | | | | | | | | | | | | | | | | | | | | | | | | | | | | | | | | | | |
| | 2 | | | | | | | | | | | | | | | | | | | | | | | | | | | | | | | | | | | | |
| | 3 | | | | | | | | | | | | | | | | | | | | | | | | | | | | | | | | | | | | |
| | 4 | | | | | | | | | | | | | | | | | | | | | | | | | | | | | | | | | | | | |
| it6 | 1 | | | | | | | | | | | | | | | | | | | | | | | | | | | | | | | | | | | | |
| | 2 | | | | | | | | | | | | | | | | | | | | | | | | | | | | | | | | | | | | |
| | 3 | | | | | | | | | | | | | | | | | | | | | | | | | | | | | | | | | | | | |
| | 4 | | | | | | | | | | | | | | | | | | | | | | | | | | | | | | | | | | | | |
| it7 | 1 | | | | | | | | | | | | | | | | | | | | | | | | | | | | | | | | | | | | |
| | 2 | | | | | | | | | | | | | | | | | | | | | | | | | | | | | | | | | | | | |
| | 3 | | | | | | | | | | | | | | | | | | | | | | | | | | | | | | | | | | | | |
| | 4 | | | | | | | | | | | | | | | | | | | | | | | | | | | | | | | | | | | | |
| it8 | 1 | | | | | | | | | | | | | | | | | | | | | | | | | | | | | | | | | | | | |
| | 2 | | | | | | | | | | | | | | | | | | | | | | | | | | | | | | | | | | | | |
| | 3 | | | | | | | | | | | | | | | | | | | | | | | | | | | | | | | | | | | | |
| | 4 | | | | | | | | | | | | | | | | | | | | | | | | | | | | | | | | | | | | |
| it9 | 1 | | | | | | | | | | | | | | | | | | | | | | | | | | | | | | | | | | | | |
| | 2 | | | | | | | | | | | | | | | | | | | | | | | | | | | | | | | | | | | | |
| | 3 | | | | | | | | | | | | | | | | | | | | | | | | | | | | | | | | | | | | |
| | 4 | | | | | | | | | | | | | | | | | | | | | | | | | | | | | | | | | | | | |
| it10 | 1 | | | | | | | | | | | | | | | | | | | | | | | | | | | | | | | | | | | | |
| | 2 | | | | | | | | | | | | | | | | | | | | | | | | | | | | | | | | | | | | |
| | 3 | | | | | | | | | | | | | | | | | | | | | | | | | | | | | | | | | | | | |
| | 4 | | | | | | | | | | | | | | | | | | | | | | | | | | | | | | | | | | | | |
| it11 | 1 | | | | | | | | | | | | | | | | | | | | | | | | | | | | | | | | | | | | |
| | 2 | | | | | | | | | | | | | | | | | | | | | | | | | | | | | | | | | | | | |
| | 3 | | | | | | | | | | | | | | | | | | | | | | | | | | | | | | | | | | | | |
| | 4 | | | | | | | | | | | | | | | | | | | | | | | | | | | | | | | | | | | | |
| it12 | 1 | | | | | | | | | | | | | | | | | | | | | | | | | | | | | | | | | | | | |
| | 2 | | | | | | | | | | | | | | | | | | | | | | | | | | | | | | | | | | | | |
| | 3 | | | | | | | | | | | | | | | | | | | | | | | | | | | | | | | | | | | | |
| | 4 | | | | | | | | | | | | | | | | | | | | | | | | | | | | | | | | | | | | |
| it13 | 1 | | | | | | | | | | | | | | | | | | | | | | | | | | | | | | | | | | | | |
| | 2 | | | | | | | | | | | | | | | | | | | | | | | | | | | | | | | | | | | | |
| | 3 | | | | | | | | | | | | | | | | | | | | | | | | | | | | | | | | | | | | |
| | 4 | | | | | | | | | | | | | | | | | | | | | | | | | | | | | | | | | | | | |
| it14 | 1 | | | | | | | | | | | | | | | | | | | | | | | | | | | | | | | | | | | | |
| | 2 | | | | | | | | | | | | | | | | | | | | | | | | | | | | | | | | | | | | |
| | 3 | | | | | | | | | | | | | | | | | | | | | | | | | | | | | | | | | | | | |
| | 4 | | | | | | | | | | | | | | | | | | | | | | | | | | | | | | | | | | | | |

| Total | Solar | 0 | 0 | 6 | 0 | 5 | 1 | 0 | 0 | 8 | 0 | 8 | 8 | 0 | 0 | 10 | 8 | 14 | 10 | 0 | 0 | 9 | 9 | 9 | 9 | 5 | 6 | 11 | 11 | 11 | 11 | 6 | 6 | 6 | 6 | 6 | 6 |
|---|---|---|---|---|---|---|---|---|---|---|---|---|---|---|---|---|---|---|---|---|---|---|---|---|---|---|---|---|---|---|---|---|---|---|---|---|---|
| | Wind | 2 | 0 | 5 | 0 | 0 | 5 | 7 | 0 | 9 | 0 | 0 | 8 | 10 | 5 | 14 | 1 | 0 | 12 | 7 | 5 | 9 | 3 | 3 | 9 | 11 | 6 | 11 | 9 | 7 | 11 | 6 | 6 | 6 | 6 | 6 | 6 |

Furthermore, two main pieces of information can be deducted from Table 1. First, the final solution and second, the decision making of the model to get there. Regarding the first one, it is interesting to see, that in all final solutions Region 3 is always selected by the model to experience a Dunkelflaute. This shows the high impact of low solar and wind availability in this region on the total system cost.

Second, by analyzing how often the models select a specific region, we can assess the severity of regional impacts. Therefore, the lower part of Table 1 presents, for each region, the sum of all low solar PV or wind events over all iterations. Here, we also find Region 3 to be chosen the most across all scenarios, followed by Region 6. This shows that lowering the renewable availability potential in these regions during this period (January) is in general hard to compensate for the system. Furthermore, the data allows to differentiate between the affected generation technologies. We can see for the ARO1 - ARO4 scenarios that Region 1 and Region 2 are affected by low wind and Region 4 by low solar PV events. This outcome reflects the construction of robust scenarios: within each region, availability reductions are first applied to the renewable technology with the highest baseline potential. For example, solar PV availability is reduced first in solar-rich regions such as Italy, while wind availability is reduced first in wind-dominated regions such as Scandinavia and the UK. The results highlight the vulnerability of regions who are overly dependent on a single dominant renewable source. A robust system configuration must therefore diversify the generation mix to mitigate this exposure, as discussed in the following sections.

## 5.2 Robust System – Cost and Investments

Figure 4 illustrates the worst-case weather realizations for each scenario, along with the resulting total system costs of the robust solution and regional cost contributions. These outcomes represent the point at which the ARO model's min-max optimization converges - where no further cost reductions are possible through technology adoption, and no higher objective values can be achieved through alternative uncertainty realizations.

The scenario-by-scenario comparison shows how increasing the uncertainty budget alters the geographic coverage of low-availability events. In ARO1 (top left of Figure 4), the model identifies low availability of both solar PV and wind in Region 3 as the worst-case scenario - effectively a worst-case regional Dunkelflaute. In ARO2 (top right), where the number of low-availability regions is increased to two, the selected regions differ from ARO1. However, Region 3 is again identified as experiencing a Dunkelflaute. Additionally, Region 6 faces low solar PV availability, while Region 1 experiences low wind. In ARO3, both Region 3 and Region 6 are affected by Dunkelflaute conditions, and Region 5 shows reduced solar PV availability. We observe that the model now selects Region 2,rather than Region 1,to experience low wind. In ARO4, low wind in Region 1 and low solar in Region 4 are added, resulting in all regions experiencing at least one type of low availability. ARO5 further includes low availability in Regions 2 and 4, and in ARO6, Dunkelflaute conditions simultaneously affect all six regions.

The lower part of Figure 4 shows the progression of total system costs (i.e., the model's objective value), which includes annual investment and generation costs as defined in Equation (1) in Subsection 4.1. It also displays the average electricity cost per scenario. In the Base scenario, total system costs amount to €151 billion, with an average electricity cost of €51/MWh. In ARO1, costs rise moderately by 9 %, suggesting that the integrated European system can efficiently absorb localized renewable scarcity events. However, as more regions experience low availability, system costs increase rapidly: in ARO2, costs reach €199 billion (+31 %) with €66/MWh on average; in ARO3, they rise to €228 billion (+50 %); in ARO4, to €245 billion (+62 %); and in ARO5, to €256 billion (+70 %) with an average cost of €85/MWh. The increase in ARO6 is marginal, reaching €258.5 billion (+71 %) and €86/MWh on average. This steep rise followed by a plateau indicates a nonlinear system response, where early regional shocks are manageable, but broader events quickly exhaust cost-effective adaptation measures.

The regional cost shares in Figure 5 further illustrate how the system cost burden is distributed. Across all scenarios, Regions 3, 5, and 6 contribute the most to total system costs. Nearly all regions, except Region 6, approximately double their costs from the Base to the ARO6 scenario due to shifts in their generation mix. However, the sensitivity to scenario severity varies by region. In the early scenarios, cost increases are driven mainly by Region 6, while other regions see only modest changes. From ARO3 onward, Region 6's costs begin to decline, while Regions 1, 3, and 5 show substantial increases. This pattern suggests that Region 6 is particularly vulnerable to smaller-scale Dunkelflaute events, where it must absorb the renewable shortfall on its own. As more regions are affected in later scenarios, the compensation burden is redistributed, alleviating pressure on Region 6 and helping stabilize its supply.

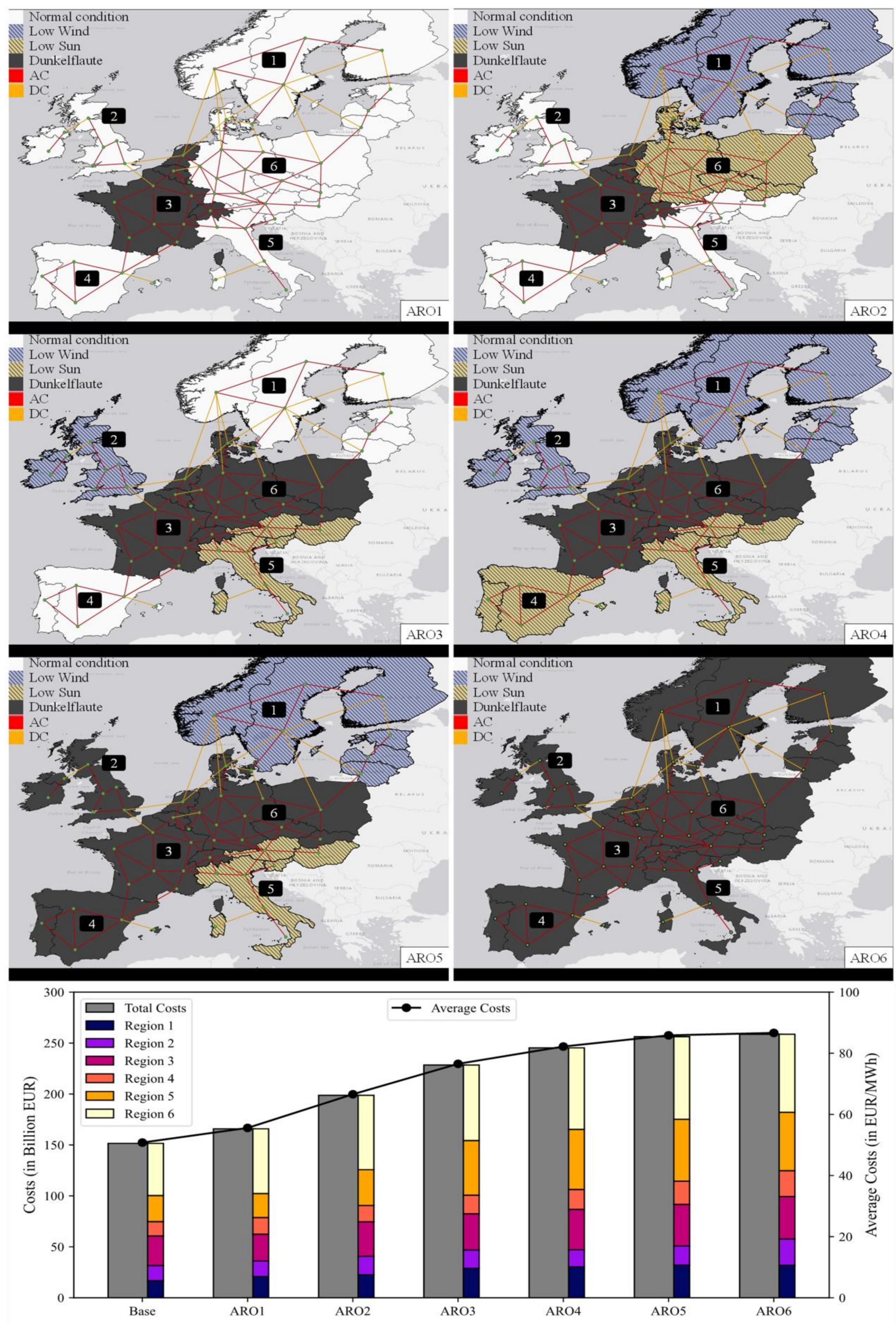

**Figure 4:** Scenario specific worst case regional weather realizations, maps for AR01, AR02, AR03, AR04, AR05, AR06, and system costs with regional shares and average electricity costs, own illustration.

Figure 5 reveals the trends in technology-specific investments. Onshore wind appears consistently across all regions, while significant solar PV deployment is concentrated in Regions 4 and 5 and, depending on the scenario, also in Regions 3 and 6. In particular, Regions 3 and 6 show more prominent solar PV investments in low-coverage scenarios (ARO1–ARO3), which decline as the number of affected regions increases in higher-coverage scenarios (ARO4–ARO5).

Additionally, we observe a substantial increase in investment cost for long-term hydrogen storage and load shedding as more regions are impacted by low availability. Hydrogen storage investments become significant starting in ARO2, especially in Regions 1, 3, 5, and 6. By ARO6, hydrogen storage accounts for 19 %–25 % of the regional system costs. Load shedding also becomes a relevant measure from ARO2 onward, with the model increasingly relying on both storage technologies and demand curtailment to meet the final megawatts of residual demand during extreme renewable scarcity events.

These findings demonstrate that the optimal robust investment mix is shaped not only by the severity of *Dunkelflaute* events but also by their regional resource endowments. In particular, long-term storage technologies emerge as critical components of the robust solution once renewable scarcity events affect a broader portion of the system.

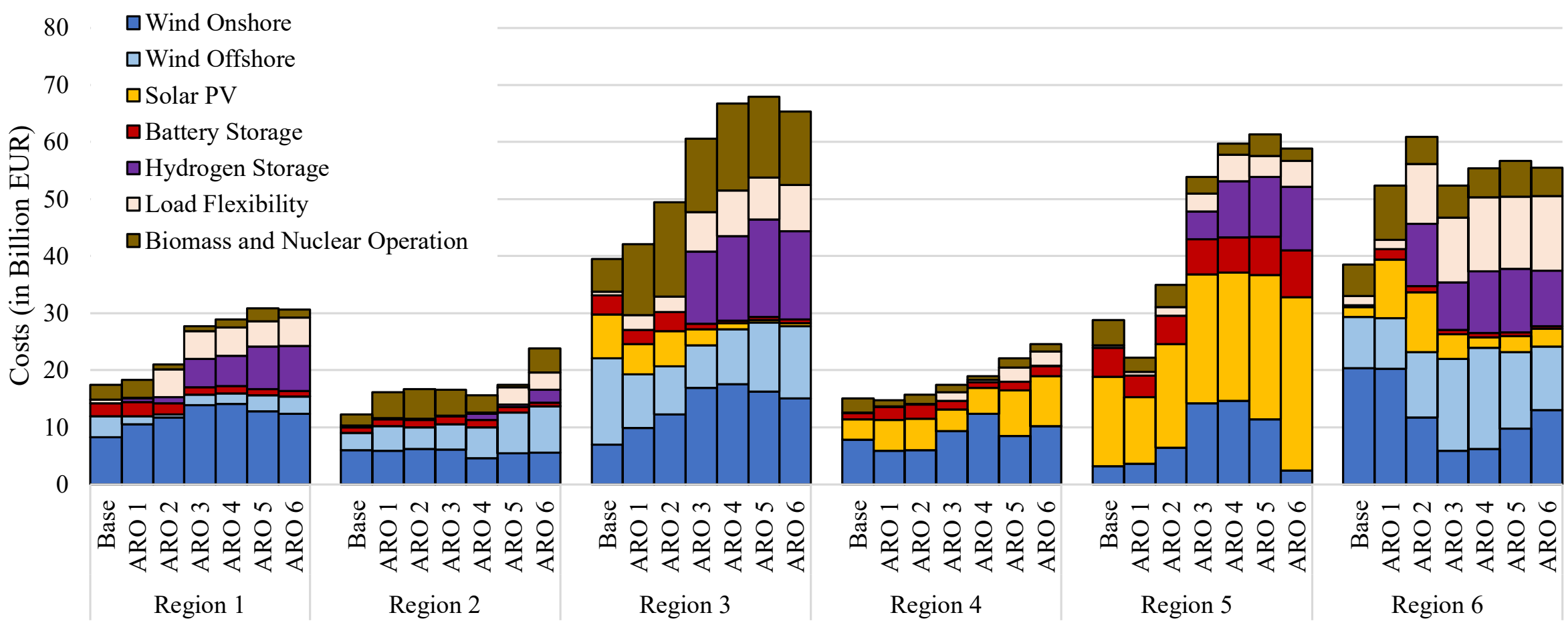

**Figure 5:** Regional and scenario-specific electricity investment, generation, and load shedding costs, own illustration.

## 5.3 Installed Capacity and Regional Generation Mix

Figure 6 depicts the scenario-specific installed capacities for each technology, aggregated across Europe. Despite its modest share in total system costs, solar PV comprises the largest share of installed capacity, reaching up to 50 % across all scenarios. Wind power represents the second-largest share, with onshore wind dominating over offshore. Battery inverters form the third-largest component of installed capacity. Investments in battery inverters increase notably when one or two regions are affected by low availability events but decline and stabilize as more regions become impacted. Overall, battery deployment closely follows the pattern of solar PV investments.

In line with the cost trend, the installed capacities of hydrogen technologies, specifically electrolyzers (for charging) and $H_2$-fired OCGTs (for discharging), become significant from ARO3 onwards, with 66 GW of electrolyzer and 45 GW of OCGT capacity. These capacities continue to grow through ARO6, reaching approximately 98 GW of electrolyzers and 70 GW of OCGTs, together accounting for around 6 % of total installed capacity. This trend reinforces the role of long-term hydrogen storage as a preferred solution for handling large-scale Dunkelflaute events.

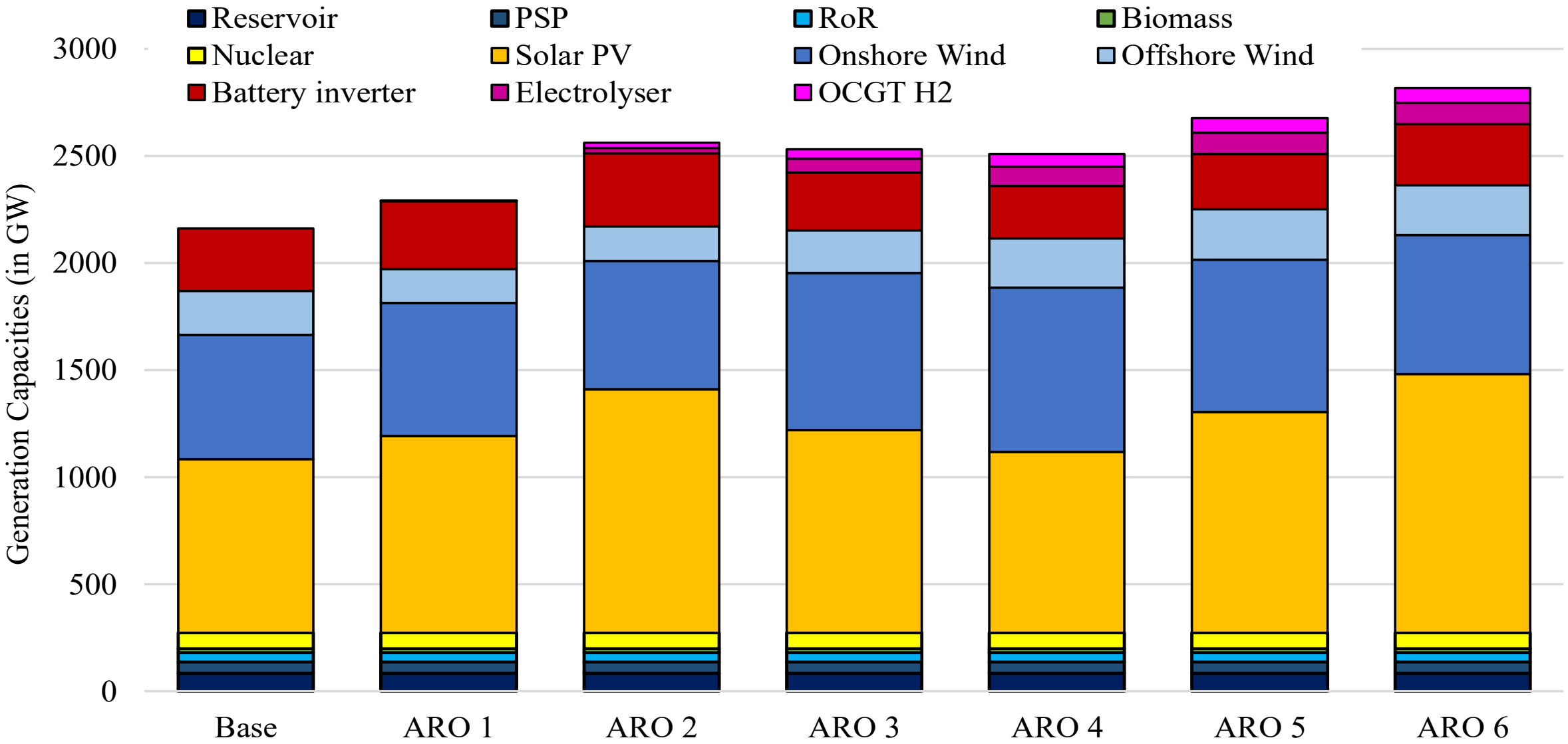

**Figure 6:** Scenario-specific technology capacity mix, own illustration.

Figure 7 presents the regionally differentiated, scenario-specific annual electricity generation (in TWh), alongside regional electricity demand indicated by red dotted lines. When a region's total generation exceeds this line, it functions as a net exporter. Electricity generation is dominated by wind power in most regions, while solar PV plays a more prominent role in Southern Europe. Most regions are either net exporters or nearly self-sufficient, with the exception of Region 6, where electricity demand significantly exceeds domestic generation. This outcome is primarily driven by the spatial mismatch between renewable resource availability and electricity demand. Region 6 has the highest electricity demand but has comparatively limited renewable generation potential. It also has the lowest share of firm, carbon-free capacity (hydropower, biomass, and nuclear) relative to its demand. Consequently, in the cost-optimal system Region 6 relies on imports from neighboring regions with more favorable renewable resources, particularly Regions 1, 3, and 5. A more detailed analysis of this aspect can be found in the Appendix E. Under extreme weather events, Region 6 also exerts a strong influence on the capacity expansion decisions in surrounding regions. For instance, while Region 3 primarily deploys offshore wind in the Base scenario, it shifts toward onshore wind under the ARO scenarios to reduce transmission distances to Region 6. When only Region 3 is affected by low availability (e.g., in ARO1), Region 6 responds with increased domestic generation investment. However, as additional regions are impacted, Region 6 increasingly relies on hydrogen storage and hydrogen-fired OCGT capacity to bridge the scarcity period. Although hydrogen technologies account for a substantial share of total system costs, their contribution to electricity generation remains modest. A similar pattern is seen with load shedding, which stays below 1 % of total generation in all scenarios. This stark contrast between cost and energy output highlights the high expense of maintaining backup capacity to safeguard the system during extreme low availability periods. Identifying cost-effective, large-scale backup solutions remains a critical challenge for reducing total system costs. Detailed regional and scenario-specific analysis is provided in Appendix E.

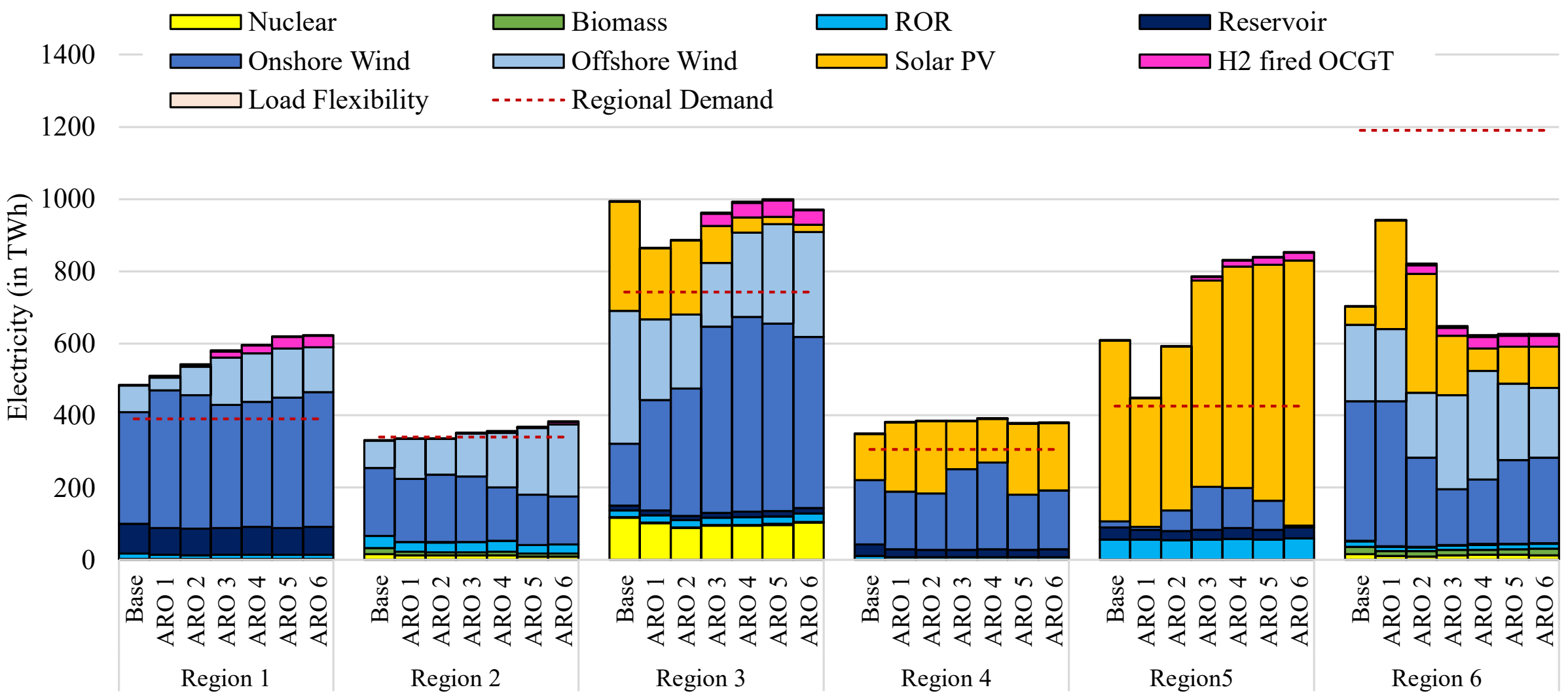

**Figure 7:** Regional generation mix in relation to the regional demand, own illustration.

## 5.4 Storage Capacity Expansion

Figure 8 presents the scenario specific regional storage capacity expansion using the storage-demand ratio on the primary as well as the total regional storage capacity on the secondary Y-axis. The storage-demand enables the comparison of storage requirements across systems and serves as a proxy for the flexibility storage provides to the energy system. For more detailed insights on this metric, the interested reader is referred to Zerrahn et al. (2018) and to Appendix F.

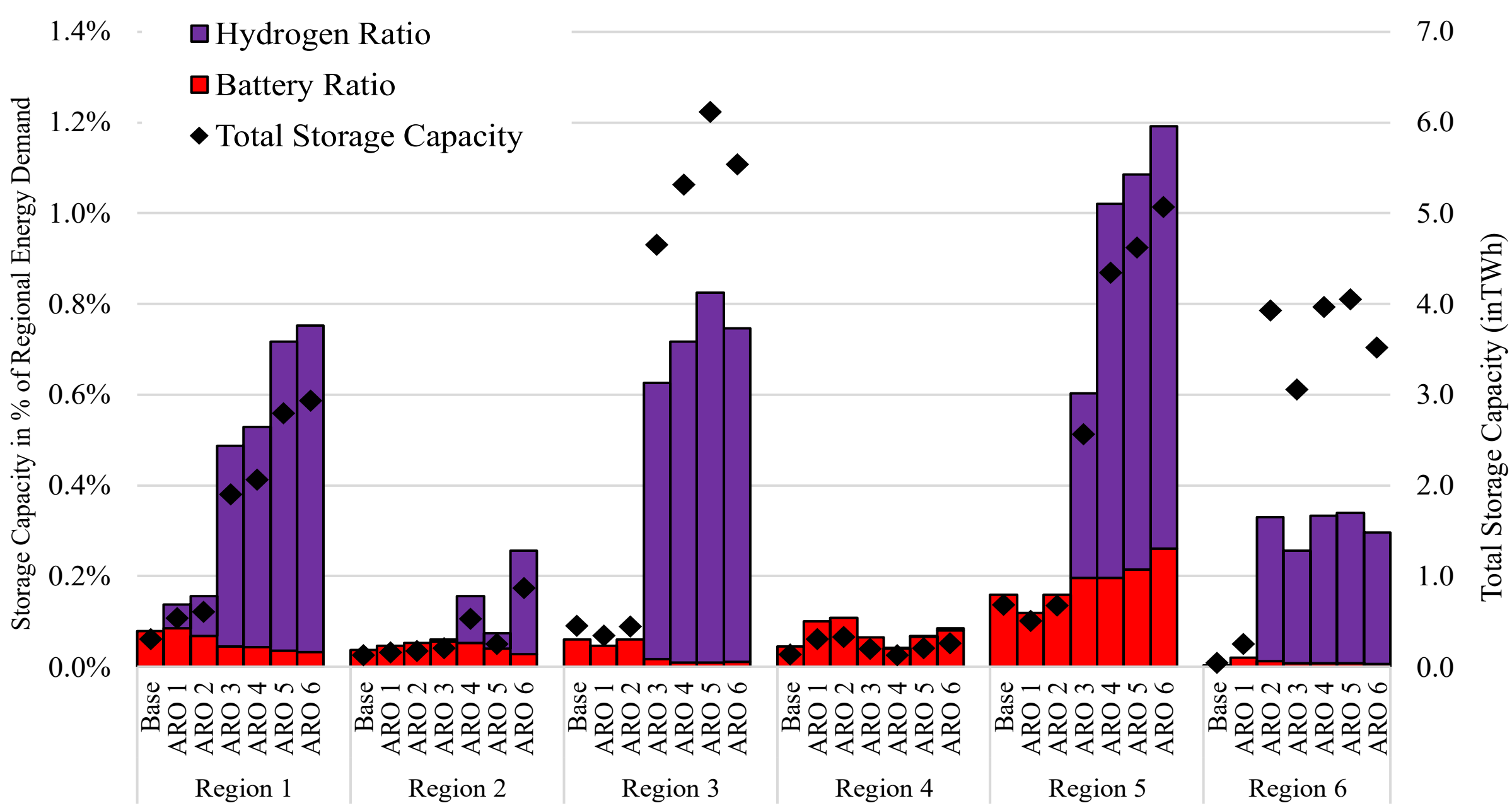

**Figure 8:** Regional storage-demand ratio plotted with stacked bars for battery storage capacity and hydrogen storage capacity on the primary axis, regional storage capacity plotted with black markers on the secondary axis, own illustration.

The figure presents two different findings: first, the absolute installed storage capacity shows that in the first two ARO scenarios most of the storage capacity in each region, but Region 6, comes from battery systems. As solar PV generation in Region 3 and Region 5 plays a major role, so does the battery storage capacity. Starting from the ARO2 scenario, the hydrogen storage expansion becomes significant, mainly driven by expansion in Region 1 and Region 6. Notably, in the ARO2 scenario, Region 6 significantly increases its hydrogen storage capacity, constructing more storage capacity (4 TWh) than all the other regions combined. However, from the ARO3 scenario Region 1, Region 4 and Region 5 are catching up, making Region 3 the region with the highest installed hydrogen storage capacity. The pronounced increase in hydrogen storage capacity between ARO2 and ARO3 may reflect a structural tipping point in the model. From ARO3 onward, four out of six regions are simultaneously affected by low renewable availability events, reducing the benefits of spatial diversification. As a result, the marginal cost of supplying demand via remote renewable generation and transmission exceeds the cost of deploying local long-duration storage, leading to a regime shift in the optimal technology portfolio. For the ARO6 scenario, the hydrogen storage capacity for the system totals 16.5 TWh and the battery storage capacity 1.7 TWh.

Second, the storage-demand ratio shows that storage expansion is decoupled from the region's energy demand. From the ARO3 scenario onward, Region 1, Region 3 and Region 5 built proportionally the most storage capacity to compensate for the renewable scarcity period, indicating that they require disproportionately more storage to cope with renewable scarcity periods. On the contrary, Region 2, Region 4 and Region 6 remain constantly below the average storage-demand ratio, showing on an overall system level a lower adoption sensibility against the modelled extreme. A regional comparison including the average storage-demand ratio across all regions is presented in Figure 14 in the Appendix F. To evaluate the role of long-term hydrogen storage in ensuring system stability, Table 2 presents the installed $H_2$-fired OCGT capacity and the maximum hydrogen storage discharge duration for the ARO6 scenario as a representative worst-case example.

**Table 2:** Maximum discharging capacity and duration of long-term storage in the ARO6 scenario.

| Region | Region 1 | Region 2 | Region 3 | Region 4 | Region 5 | Region 6 |
|---|---|---|---|---|---|---|
| $H_2$ fired OCGT in GW | 11.8 | 4.0 | 24.1 | 0.1 | 16.5 | 13.9 |
| Duration in hours | 139 | 115 | 132 | 112 | 142 | 147 |

The table indicates that hydrogen storage is dimensioned to cover between four and six days of demand, aligning closely with the modeled seven-day renewable scarcity event duration. The capacities of both hydrogen storage and OCGT plants are optimized accordingly. Region 3 and Region 5 install the highest levels of OCGT capacity, enabling them to support not only their own loads but to also partially cover Region 6 during a renewable scarcity period. This finding is evaluated in more detail in Appendix F

These results highlight how worst-case scarcity events identified by the model shape the regional demand for long-duration storage, revealing the cost sensitivity of different regions under stress. The earlier hydrogen storage appears in the optimal capacity mix, the more exposed a region is to extreme events - suggesting that lower-cost options such as load flexibility, transmission, or renewable generation expansion have already been exhausted. An alternative would be to expand other firm capacity like oil power plants or carbon-neutral nuclear power plants to cover the period, which remains outside the scope of this analysis.

# 6 DISCUSSION AND CONCLUSION

This study introduces the first adaptive robust capacity expansion model that endogenously incorporates worst-case weather events into the long-term planning of a decarbonized electricity system spanning 24 European countries. A key advantage of the ARO framework lies in its sequential decision-making structure, which allows the model to anticipate and respond to the most severe realizations of weather-related uncertainty. By treating periods of low solar and wind availability as worst-case events, the model identifies their potential timing and regional occurrence, quantifies their system-wide impact, and determines the adjustments required to ensure a robust and resilient technology mix.

From a methodological perspective, this study advances the literature on robust European energy systems by introducing an optimization framework that captures the regional distribution of weather-related uncertainty and system adaptation. While existing studies assess the overall system-wide effects of weather robustness, our approach goes further by distinguishing between regional impacts related to capacity adoptions and cost developments. On this basis our model reveals region-specific adaptation strategies and vulnerabilities, emphasizing the importance of incorporating regional differentiation into robust energy system planning.

Our results reveal on a system wide (European) basis a nonlinear cost response as the geographical scope of low availability events increases. Compared to the baseline scenario (€151.6 billion), system costs rise moderately under localized shocks (+9% in ARO1), but increase sharply when broader impacts are considered (+31% in ARO2, +51% in ARO3). As more regions are affected, marginal cost increases taper off: +62 % in ARO4, +69 % in ARO5, and +71 % in ARO6. The nonlinear shape of the cost–adaptation curve reflects how the system's ability to balance renewable scarcity events evolves as the spatial extent of these events increases. Under localized shocks (ARO1), cost increases remain modest because unaffected regions can effectively compensate through cross-border energy exchanges, requiring only limited additional storage or renewable capacity. However, as the geographical scope widens (ARO2 and ARO3), regions that previously served as balancing partners are themselves affected, sharply reducing the system's inherent flexibility. This triggers a disproportionately large rise in system costs, driven by the need to expand storage and renewable generation to shift surplus energy into scarcity periods. Once most regions are already impacted (ARO4–ARO6), the cost curve begins to flatten: much of the necessary long-duration storage and excess generation has already been built, and the remaining shortfalls can be compensated more easily within the newly affected regions themselves. As a result, marginal adaptation costs decline, and the curve approaches a saturation point.

Comparison with previous studies shows that ARO1 aligns closely with deterministic models such as Gøtske et al. (2024) and Grochowicz et al. (2024) while ARO2 yields results similar to those of Plaga and Bertsch (2022). The significantly higher costs observed in ARO3 and beyond likely stem from our model design, which combines multiple worst-case events and enforces a strict carbon cap, excluding other low-carbon backup options such as nuclear power, and incorporates relatively high costs for load shedding. While these assumptions may yield conservative estimates, they enable us to explore the upper bounds of system cost escalation and to identify the technologies required to hedge against large-scale Dunkelflaute events across Europe. Furthermore, the results underscore the strong sensitivity of system configurations to the geographic scope of such extreme events. This emphasizes the need for system operators and policymakers to develop a clear understanding of which types of events to anticipate. Consequently, harmonizing assumptions regarding extreme weather scenarios is a key prerequisite for planning a weather-resilient European energy system.

At the regional level, our results provide insights into which regions are mostly affected by worst-case renewable scarcity events and how the cost increases associated with a broader geographic extent of these events are distributed across regions. We identify which regions contribute most to overall system costs and demonstrate how the burden of adaptation shifts—from central regions under localized events to more peripheral regions as the spatial coverage of the scarcity event expands.

The ARO model consistently identifies central European regions, particularly Regions 3 and 6, as critical. Region 6, which encompasses major demand centers like Germany, relies on imports from Region 3, a strong renewable producer. When either is affected by low availability, the cost of reconfiguring the system increases substantially. In contrast, peripheral regions (e.g., Regions 1, 2, 4, and 5) are less frequently identified as critical extreme event locations but still face high adaptation costs when central regions are impacted. This is particularly evident for Regions 1 and Region 5, which disproportionately expand their storage infrastructure to help maintain overall system balance. Beginning with the ARO-3 scenario, both regions consistently exhibit a storage-to-demand ratio well above the system-wide average, whereas other regions, most notably Region 6, show ratios below the average. These findings underscore the practical relevance of a coordinated European energy policy. While central European countries benefit from a system-wide optimization perspective, peripheral countries may face higher system cost. Ongoing debates in countries such as Norway[5] and Sweden[6] - where high electricity prices have been attributed to exports during periods of renewable scarcity, particularly to neighboring countries like Denmark and Germany - illustrate the emerging tensions. These developments serve as an early indication of the distributional challenges that may intensify as Europe transitions toward a fully renewable electricity system. To avoid future fragmentation and resource nationalism driven by national priorities, our work demonstrates that a robust system on a European scale might be cost efficient in handling extreme renewable scarcity events, while on a local or regional scale the cost burdens are unevenly distributed. To promote social and political acceptance policymakers should not only incentivize investments in renewables, transmission infrastructure, and storage in high-potential regions, but also consider cross-border cost compensation or benefit-sharing mechanisms.

Another key insight is the shift in optimal system configuration as the geographical scale of the renewable scarcity event grows. For small-scale events (ARO1, ARO2), expanding renewables, batteries, and transmission capacity is cost-effective. As interregional balancing becomes constrained by widespread weather events, long-term hydrogen storage becomes essential. Although hydrogen accounts for a large share of system costs, its actual contribution to electricity generation remains modest. Likewise, load shedding remains under 1 % of total generation in all scenarios, while still significantly driving system costs. This aspect highlights the need for affordable, large-scale backup capacity.

A limitation of this study lies in the representation of weather data. First, we use a synthetic average weather year rather than a historical one for the Base scenario. While this approach captures representative renewable generation patterns, it may not fully reflect the variability and temporal extremes found in real-world data. Second, to reduce computational complexity, we apply a time series reduction technique using a four-hour moving average. This smoothing may underestimate the intensity and frequency of short-lived extreme events. Third, the model assumes static, predefined weather regions for the realization of extreme events, which does not reflect the continuous and dynamic nature of real weather systems. In reality, low availability conditions can span irregular and shifting geographical areas. Fourth, the model focuses exclusively on seven-day periods of low solar and wind availability, with uncertainty realizations restricted to a four-week window in January. As a result, extreme events occurring in other months or with different durations are not captured. These assumptions may lead to an underestimation of required year-round storage requirement, especially for seasonal storage systems.

Additional model limitations include fixed capacities for firm generation technologies (e.g., nuclear), which are based on current levels or policy targets, and constraints on transmission expansion. Allowing these capacities to expand endogenously or without limits could lead to different investment strategies. We conducted a small sensitivity analysis assuming unlimited transmission line expansion for the base case and ARO6, as well as different load-shedding costs for the ARO6 scenario. These results are presented in the supplementary materials. We also do not account for the hydrogen sector, omitting the additional system flexibility of the hydrogen transportation and storage infrastructure. This may lead to an overestimation of the modelled extreme events on the overall system costs. Finally, our results are subject to the 2050 technology cost projections. Different cost developments may lead to other optimal system configurations.

Future research could extend this framework by refining the representation of weather uncertainty. In particular, uncertainty sets could be expanded beyond binary availability reductions by introducing severity-weighted uncertainty budgets, allowing regions to experience different intensities of renewable scarcity. This would enable a more continuous representation of extreme events while preserving the tractability of robust optimization. In addition, uncertainty sets that consider correlation could be employed to reflect the spatial coherence of large-scale meteorological systems, such as continent-wide wind lulls or persistent cloud cover, rather than assuming independent regional realizations. Further extensions could relax the assumption of fixed weather regions by allowing dynamically evolving scarcity patterns that shift across space and time, better capturing the propagation of real weather systems.

[5] https://www.euronews.com/business/2024/12/13/norway-aims-to-cut-energy-links-with-europe-due-to-soaring-prices
[6] https://harici.com.tr/en/sweden-blames-germanys-nuclear-phase-out-for-energy-crisis/

Moreover, broadening the temporal scope of uncertainty realizations to include different seasons and event durations would allow the analysis of seasonal and multi-week scarcity events, improving the assessment of long-term and seasonal storage requirements.

# 7 ACKNOWLEDGEMENTS

Maximilian Bernecker gratefully acknowledges supported by the Federal Ministry of Education and Research, Award No. 19FS2032C, as well as the German Federal Government, the Federal Ministry of Education and Research, and the State of Brandenburg within the framework of the joint project EIZ: Energy Innovation Center (project numbers 85056897 and 03SF0693A) with funds from the Structural Development Act (Strukturstärkungsgesetz) for coal-mining regions."

Felix Müsgens gratefully acknowledges financial support from the Federal Ministry of Education and Research of Germany in the Ariadne project (03SFK5S0).

Iegor Riepin acknowledges support by the German Federal Ministry for Economic Affairs and Energy (BMWE) under Grant No. 03EI4083A (RESILIENT) jointly with the CETPartnership through the Joint Call 2022. As such, I.R. further acknowledges funding from the European Union's Horizon Europe research and innovation programme under grant agreement no. 101069750.

# 8 OPEN DATA STATEMENT

The model is formulated using the General Algebraic Modeling Language (GAMS). The code and data are available at github: https://github.com/bernemax/ARO_Dunkelflaute_Europe

# 9 REFERENCES

Baringo, L., Boffino, L., Oggioni, G., 2020. Robust expansion planning of a distribution system with electric vehicles, storage and renewable units. Applied Energy 265, 114679. https://doi.org/10.1016/j.apenergy.2020.114679

Baringo, L., Rahimiyan, M., 2020. Virtual Power Plants and Electricity Markets: Decision Making Under Uncertainty. Springer International Publishing, Cham. https://doi.org/10.1007/978-3-030-47602-1

Beck-Bang, A.C., Overgaard, S.B., Keles, D., 2026. Electricity supply and demand in the European market: the effect on generation adequacy in a dunkelflaute period. Energy Policy 208, 114910. https://doi.org/10.1016/j.enpol.2025.114910

Ben-Tal, A., El Ghaoui, L., Nemirovski, A., 2009. Robust Optimization | Princeton University Press [WWW Document]. URL https://press.princeton.edu/books/hardcover/9780691143682/robust-optimization (accessed 2.10.26).

Bertsimas, D., Litvinov, E., Sun, X.A., Zhao, J., Zheng, T., 2013. Adaptive Robust Optimization for the Security Constrained Unit Commitment Problem. IEEE Trans. Power Syst. 28, 52–63. https://doi.org/10.1109/TPWRS.2012.2205021

Bertsimas, D., Shtern, S., Sturt, B., 2020. Two-stage sample robust optimization. https://doi.org/10.48550/arXiv.1907.07142

Bertsimas, D., Sim, M., 2004. The Price of Robustness. Operations Research 52, 35–53. https://doi.org/10.1287/opre.1030.0065

Birge, J.R., Louveaux, F., 2011. Introduction to Stochastic Programming, Springer Series in Operations Research and Financial Engineering. Springer New York, New York, NY. https://doi.org/10.1007/978-1-4614-0237-4

Bloomfield, Hannah C., Brayshaw, D.J., Charlton-Perez, A.J., 2020. Characterizing the winter meteorological drivers of the European electricity system using targeted circulation types. Meteorological Applications 27, e1858. https://doi.org/10.1002/met.1858

Bloomfield, H. C., Suitters, C.C., Drew, D.R., 2020. Meteorological Drivers of European Power System Stress. Journal of Renewable Energy 2020, e5481010. https://doi.org/10.1155/2020/5481010

Brown, T., Schlachtberger, D., Kies, A., Schramm, S., Greiner, M., 2018. Synergies of sector coupling and transmission reinforcement in a cost-optimised, highly renewable European energy system. Energy 160, 720–739. https://doi.org/10.1016/j.energy.2018.06.222

Caglayan, D.G., Heinrichs, H.U., Robinius, M., Stolten, D., 2021. Robust design of a future 100% renewable european energy supply system with hydrogen infrastructure. International Journal of Hydrogen Energy, HYDROGEN ENERGY SYSTEMS 46, 29376–29390. https://doi.org/10.1016/j.ijhydene.2020.12.197

Conejo, A.J., Baringo Morales, L., Kazempour, S.J., Siddiqui, A.S., 2016. Investment in Electricity Generation and Transmission. Springer International Publishing, Cham. https://doi.org/10.1007/978-3-319-29501-5

Davis, S.J., Lewis, N.S., Shaner, M., Aggarwal, S., Arent, D., Azevedo, I.L., Benson, S.M., Bradley, T., Brouwer, J., Chiang, Y.-M., Clack, C.T.M., Cohen, A., Doig, S., Edmonds, J., Fennell, P., Field, C.B., Hannegan, B., Hodge, B.-M., Hoffert, M.I., Ingersoll, E., Jaramillo, P., Lackner, K.S., Mach, K.J., Mastrandrea, M., Ogden, J., Peterson, P.F., Sanchez, D.L., Sperling, D., Stagner, J., Trancik, J.E., Yang, C.-J., Caldeira, K., 2018. Net-zero emissions energy systems. Science 360, eaas9793. https://doi.org/10.1126/science.aas9793

Gøtske, E.K., Andresen, G.B., Neumann, F., Victoria, M., 2024. Designing a sector-coupled European energy system robust to 60 years of historical weather data. Nat Commun 15, 10680. https://doi.org/10.1038/s41467-024-54853-3

Grochowicz, A., Greevenbroek, K. van, Benth, F.E., Zeyringer, M., 2023. Intersecting near-optimal spaces: European power systems with more resilience to weather variability. Energy Economics 106496. https://doi.org/10.1016/j.eneco.2022.106496

Grochowicz, A., van Greevenbroek, K., Bloomfield, H.C., 2024. Using power system modelling outputs to identify weather-induced extreme events in highly renewable systems. Environ. Res. Lett. 19, 054038. https://doi.org/10.1088/1748-9326/ad374a

Jabr, R.A., 2013. Robust Transmission Network Expansion Planning With Uncertain Renewable Generation and Loads. IEEE Trans. Power Syst. 28, 4558–4567. https://doi.org/10.1109/TPWRS.2013.2267058

Jurasz, J., Mikulik, J., Dąbek, P.B., Guezgouz, M., Kaźmierczak, B., 2021. Complementarity and 'Resource Droughts' of Solar and Wind Energy in Poland: An ERA5-Based Analysis. Energies 14, 1118. https://doi.org/10.3390/en14041118

Li, B., Basu, S., Watson, S.J., Russchenberg, H.W.J., 2021a. A Brief Climatology of Dunkelflaute Events over and Surrounding the North and Baltic Sea Areas. Energies 14, 6508. https://doi.org/10.3390/en14206508

Li, B., Basu, S., Watson, S.J., Russchenberg, H.W.J., 2021b. Mesoscale modeling of a "Dunkelflaute" event. Wind Energy 24, 5–23. https://doi.org/10.1002/we.2554

Li, B., Basu, S., Watson, S.J., Russchenberg, H.W.J., 2020. Quantifying the Predictability of a 'Dunkelflaute' Event by Utilizing a Mesoscale Model. J. Phys.: Conf. Ser. 1618, 062042. https://doi.org/10.1088/1742-6596/1618/6/062042

Mattsson, N., Verendel, V., Hedenus, F., Reichenberg, L., 2021. An autopilot for energy models – Automatic generation of renewable supply curves, hourly capacity factors and hourly synthetic electricity demand for arbitrary world regions. Energy Strategy Reviews 33, 100606. https://doi.org/10.1016/j.esr.2020.100606

Mayer, M.J., Biró, B., Szücs, B., Aszódi, A., 2023. Probabilistic modeling of future electricity systems with high renewable energy penetration using machine learning. Applied Energy 336, 120801. https://doi.org/10.1016/j.apenergy.2023.120801

Mínguez, R., García-Bertrand, R., 2016. Robust transmission network expansion planning in energy systems: Improving computational performance. European Journal of Operational Research 248, 21–32. https://doi.org/10.1016/j.ejor.2015.06.068

Ohlendorf, N., Schill, W.-P., 2020. Frequency and duration of low-wind-power events in Germany. Environ. Res. Lett. 15, 084045. https://doi.org/10.1088/1748-9326/ab91e9

Otero, N., Martius, O., Allen, S., Bloomfield, H., Schaefli, B., 2022a. Characterizing renewable energy compound events across Europe using a logistic regression-based approach. Meteorological Applications 29, e2089. https://doi.org/10.1002/met.2089

Otero, N., Martius, O., Allen, S., Bloomfield, H., Schaefli, B., 2022b. A copula-based assessment of renewable energy droughts across Europe. Renewable Energy 201, 667–677. https://doi.org/10.1016/j.renene.2022.10.091

Perera, A.T.D., Nik, V.M., Chen, D., Scartezzini, J.L., Hong, T., 2020. Quantifying the impacts of climate change and extreme climate events on energy systems. Nature Energy 5, 150–159. https://doi.org/10.1038/s41560-020-0558-0

Plaga, L.S., Bertsch, V., 2022. Robust planning of a European Electricity System under climate uncertainty, in: 2022 18th International Conference on the European Energy Market (EEM). Presented at the 2022 18th International Conference on the European Energy Market (EEM), pp. 1–8. https://doi.org/10.1109/EEM54602.2022.9921057

Raynaud, D., Hingray, B., François, B., Creutin, J.D., 2018. Energy droughts from variable renewable energy sources in European climates. Renewable Energy 125, 578–589. https://doi.org/10.1016/j.renene.2018.02.130

Reichenberg, L., Hedenus, F., Odenberger, M., Johnsson, F., 2018. The marginal system LCOE of variable renewables – Evaluating high penetration levels of wind and solar in Europe. Energy 152, 914–924. https://doi.org/10.1016/j.energy.2018.02.061

Riepin, I., Schmidt, M., Baringo, L., Müsgens, F., 2022. Adaptive robust optimization for European strategic gas infrastructure planning. Applied Energy 324, 119686. https://doi.org/10.1016/j.apenergy.2022.119686

Roald, L.A., Pozo, D., Papavasiliou, A., Molzahn, D.K., Kazempour, J., Conejo, A., 2023. Power systems optimization under uncertainty: A review of methods and applications. Electric Power Systems Research 214, 108725. https://doi.org/10.1016/j.epsr.2022.108725

Rogelj, J., Luderer, G., Pietzcker, R.C., Kriegler, E., Schaeffer, M., Krey, V., Riahi, K., 2015. Energy system transformations for limiting end-of-century warming to below 1.5 °C. Nature Clim Change 5, 519–527. https://doi.org/10.1038/nclimate2572

Roldán, C., García-Bertrand, R., Mínguez, R., 2020. Robust transmission expansion planning with uncertain generations and loads using full probabilistic information. Electric Power Systems Research 189, 106793. https://doi.org/10.1016/j.epsr.2020.106793

Roldan, C., Minguez, R., Garcia-Bertrand, R., Arroyo, J.M., 2019. Robust Transmission Network Expansion Planning Under Correlated Uncertainty. IEEE Trans. Power Syst. 34, 2071–2082. https://doi.org/10.1109/TPWRS.2018.2889032

Roldán, C., Sánchez de la Nieta, A.A., García-Bertrand, R., Mínguez, R., 2018. Robust dynamic transmission and renewable generation expansion planning: Walking towards sustainable systems. International Journal of Electrical Power & Energy Systems 96, 52–63. https://doi.org/10.1016/j.ijepes.2017.09.021

Ruhnau, O., Qvist, S., 2022. Storage requirements in a 100% renewable electricity system: extreme events and inter-annual variability. Environmental Research Letters 17, 044018. https://doi.org/10.1088/1748-9326/ac4dc8

Schlachtberger, D.P., Brown, T., Schramm, S., Greiner, M., 2017. The benefits of cooperation in a highly renewable European electricity network. Energy 134, 469–481. https://doi.org/10.1016/j.energy.2017.06.004

Zerrahn, A., Schill, W.-P., Kemfert, C., 2018. On the economics of electrical storage for variable renewable energy sources. European Economic Review 108, 259–279. https://doi.org/10.1016/j.euroecorev.2018.07.004

Zhang, X., Conejo, A.J., 2018. Robust Transmission Expansion Planning Representing Long- and Short-Term Uncertainty. IEEE Trans. Power Syst. 33, 1329–1338. https://doi.org/10.1109/TPWRS.2017.2717944

# APPENDIX A

### Expansion Planning and Adaptive Robust Optimization

In general, expansion problems address the need to plan how future demand can be met with scarce resources, and they aim to either maximize or minimize the objective criteria under certain technical constraints. In the context of the electrical energy system, the aim is to identify how future energy needs can be met, investing in generation assets and transmission lines to meet the demand (Conejo et al., 2016). A typical optimization problem from a central planner's perspective could look like the following Equation A1:

$$\begin{aligned} & Min_x \;\; IC^T x \\ \text{s.t.} \quad & \\ & h(x) = 0 \\ & g(x) \leq 0 \end{aligned} \tag{A1}$$

The objective function in Eq. A1 minimizes the total system costs, which are the product of a vector x, representing an investment decision and the investment cost vector $IC^T$. In addition, certain equality conditions $h(x)$, such as supply needs to meet demand, and inequality conditions $g(x)$, such as that generation is limited to nominal capacity, must be satisfied. In this deterministic framework, uncertainty is incorporated exogenously by changing uncertain parameters and formulating deterministic scenarios, analyzing the results ex-post. In contrast, Adaptive Robust Optimization internalizes uncertainty within the optimization problem itself. The objective is to identify solutions that remain feasible and cost-efficient under worst-case realizations of uncertain parameters within predefined uncertainty sets. ARO explicitly models recourse decisions, allowing the system to adapt after uncertainty is realized. By constraining uncertainty through a defined uncertainty budget, the approach results in a three-level optimization problem, which can be formulated as presented in Eq. A2:

$$\begin{array}{lll} \text{(i) } Min_x \;\; IC^T x & \text{(ii) } Max_u & \text{(iii) } Min_y \; [OC(x,u)]^T \, y \\ \text{s.t.} & \text{s.t.} & \text{s.t.} \\ h(x) = 0 & u \in U & y \in \Omega(x,u) = \{ \\ g(x) \leq 0 & & A(x,u)y = b(x,u) \text{: } \lambda \\ & & D(x,u)y \geq e(x,u) \text{: } \mu \} \end{array} \tag{A2}$$

Here, the deterministic problem (i) represents solely the first stage of the optimization process, thus minimizing the system costs prior to the uncertainty realization. Based on this solution, the second level (ii) aims to maximize the total system cost by finding the worst-case realizations, represented by the variable vector $u$. The second-stage decision variable is bounded by the confidence bounds of the possible uncertainty realization space $U$, which could be, for example, the "on" or "off" status of a power plant. The third level encompasses the corrective actions, aiming to minimize total system operation costs $OC$ based on the prior results $(x, u)$ of level (i) and (ii). These actions are represented by variable vector $y$, and bounded by the feasibility space $\Omega(x, u)$ which is the constraining level (iii) based on (i) and (ii). Consequently, $A, b, D, e$ are matrices with constant parameters and $\lambda$ and $\mu$ represent the dual variables associated with equality and inequality constraints. In the context of the electricity energy system model, these actions could for example correspond to technical constraints of power plants in the dispatch process. This problem structure enables the model to adapt to uncertainty, following the typical chronological order of (i) decision making, (ii) uncertainty realization, and (iii) reaction in an infrastructure planning process (Ruiz and Conejo, 2015).

### Modeling Uncertainty

To model uncertainty in the second-level problem (ii), we employ polyhedral uncertainty sets, introduced by Bertsimas and Sim (2004). More specifically, we use a cardinality-constrained polyhedral uncertainty set, which allows us to adjust the level of robustness, by controlling the number of uncertain coefficient perturbations within a constraint. This number is represented by the so-called uncertainty budget $\Gamma$, which guarantees that the solution of the model is protected against all cases up to $\Gamma$ coefficients that can deviate from their nominal value.

In our second-level problem, the uncertainty is represented by the decision variable $u$, which can take values within specified confidence bounds as shown in Eq. A3:

$$u \in \{\bar{u} - \hat{u}, \bar{u} + \hat{u}\} \tag{A3}$$

Here, $\bar{u}$ and $\hat{u}$ are specifying the expected nominal and the maximum deviation value, respectively. We follow the definition of the cardinality-constrained uncertainty set Ω as described by Mínguez and García-Bertrand (2016) and Baringo et al. (2020), stated by Eq. A4:

$$\Omega=\left\{\begin{array}{c} u=\bar{u}+diag(z^{+})\,\hat{u}-diag(z^{-})\,\hat{u}, \\ z^{+},z^{-}\in\{0,1\}^{n}, \\ \sum_{k=1}^{n}(z^{+}+z^{-})\leq\Gamma \\ z^{+}+z^{-}\leq 1 \end{array}\right\} \quad \text{(A4)}$$

Here, the value of $\tilde{u}$ in depends on the binary variables $z^{+}, z^{-}$, which decide whether $u$ deviates from its nominal value towards the upper or lower bound, or not. To control the level of robustness, $\Gamma$ is restricting the number of deviations. A value of $\Gamma$=0 leads to the realization of the expected value $u=\bar{u}$, while any other discrete positive value of $\Gamma$ implies that $u$ can deviate from the expected value, which represents an uncertainty realization. Hence, $u$ cannot take values of the upper and lower bounds at the same time, and the simultaneous realization of $z^{+}$ and $z^{-}$ is excluded.

### Solution Strategy

The solution strategy on solving the three-level optimization problem is based on the literature stream established by Bertsimas et al. (2013), Mínguez and García-Bertrand (2016), Zhang and Conejo (2018), and Baringo et al. (2020). The first step is to couple the second level (ii) and the third level (iii) of the optimization problem, making it a single optimization problem. The second step is to linearize the new second level, and employ a column and constraint algorithm to solve the remaining two-level optimization problem in an efficient manner as visualized by Figure 9 below.

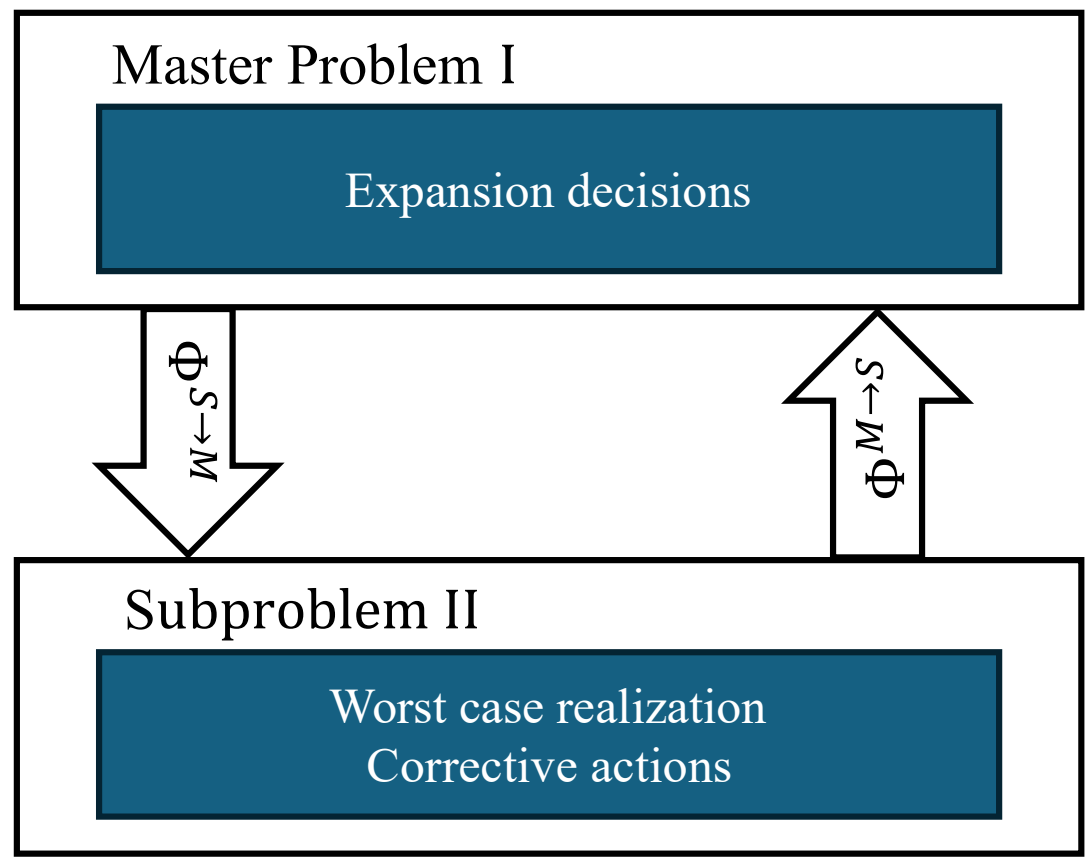

**Figure 9.** Iterative ARO solution approach, based on the work of Baringo et al. (2020).

In the following, we provide a short explanation of these processes. However, the interested reader is also referred to the work of Baringo and Rahimiyan (2020) and Riepin et al. (2022) which provide a more detailed explanation. In order to couple the third level (iii) with the second level (ii) optimization problem, we assume that problem (iii) is linear, and thus convex. On this basis we know that the dual problem solution equals the primary problem (strong duality). The reformulation of the third-level problem into its dual pendant (iii') is shown in the following (A5):

$$\text{(iii)}\ Min_{y}\ [OC(x,u)]^{T}\ y \qquad\qquad \text{(iii')}\ Max_{\lambda,\mu}\ B(x,u)\cdot\lambda+E(x,u)\cdot\mu$$

$$\text{s.t.} \qquad\qquad\qquad\qquad \text{s.t.}$$

$$y\in\Omega(x,u)=\ \{ \qquad\qquad A(x,u)\cdot\lambda+D(x,u)\cdot\mu=OC(x,u) \quad \text{(A5)}$$

$$A(x,u)y=b(x,u):\lambda \qquad\qquad \lambda: free,\ \mu\geq 0$$

$$D(x,u)y\geq e(x,u):\mu\ \}$$

As a next step, the dual reformulated problem (iii') is merged with level (ii), resulting in a single optimization problem (II) described in Eq A6. The new two-level min (I) – max (II) optimization problem is as stated below:

$$\text{(I) } Min_x \; IC^T x \qquad \text{s.t.} \qquad h(x) = 0 \qquad g(x) \leq 0$$

$$\text{(II) } Max_{u,\lambda,\mu} \; [b(x,u)\,]^T \lambda + [D(x,u)]^T \mu \qquad \text{s.t.} \qquad A(x,u)\lambda + D(x,u)\mu = OC(x,u) \qquad u \in U,\ \lambda : free,\ \mu \geq 0 \qquad \text{(A6)}$$

As there is a bilinear term $[b(x,u)\,]^T \lambda$ in the objective function of the subproblem (II), it makes this formulation a Mixed-Integer Nonlinear Programming (MINLP) and thus is hard to solve. Therefore, big-M constraints can be employed to reformulate this term, adding an auxiliary variable, making the problem a Mixed-Integer Linear Programming (MILP), which is easier to solve.

# APPENDIX B

### Table 3: Load shedding costs

| Load shedding | | | |
|---|---|---|---|
| Parameter | Value | Parameter | Value |
| $f^{L1}$ | 5% | $sc_n^{LS1}$ | 1.000 EUR/MWh |
| $f^{L2}$ | 15% | $sc_n^{LS2}$ | 3.000 EUR/MWh |
| $f^{L3}$ | 80% | $sc_n^{LS3}$ | 12.000 EUR/MWh (1) |

(1): Note that such a high load shedding level may represent an unrealistic assumption. We conducted a sensitivity analysis using the lower values provided in the Supplementary Materials.

### Table 4: Technology data

| Technology | Lifetime in years | Interest Rate | Overnight costs | Fixed OM costs | Fuel Costs | Efficiency | Data Source |
|---|---|---|---|---|---|---|---|
| Onshore Wind | 30 | 7.5% | 963 EUR/ kW | 9.63 EUR/kW/y | - | - | Danish Energy Agency |
| Offshore Wind | 30 | 9.3% | 1380 EUR/ kW | 13.80 EUR/kW/y | - | - | Danish Energy Agency |
| Solar PV | 35 | 6.9% | 370 EUR/ kW | 7.40 EUR/kW/y | - | - | Danish Energy Agency |
| Battery Inverter | 10 | 6.0% | 60.0 EUR/ kW | 0.6 EUR/kW/y | - | 96% | Danish Energy Agency |
| Battery Storage | 30 | 6.0% | 75 EUR/ kWh | 0.6 EUR/kWh/y | - | - | Danish Energy Agency |
| Electrolyzer | 25 | 8.0% | 350 EUR/ kW | 14 EUR/kW/y | - | 70% | Danish Energy Agency |
| $H_2$ Storage Tank | 30 | 8.0% | 21 EUR/ kWh | 0.5 EUR/kWh/y | - | - | Danish Energy Agency |
| $H_2$-fired OCGT | 25 | 8.0% | 411 EUR/ kW | 8.7 EUR/kW/y | - | 43% | Danish Energy Agency |
| AC Transmission | 40 | 6.0% | 700-2500 EUR/MW/km | - | - | - | NEP 2023 (1)- |
| DC Transmission | 40 | 6.0% | 2500-5500 EUR/MW/km | - | - | - | NEP 2023 (1)- |
| Biomass | 60 | - | - | 80 EUR/kW/y | 4.5 EUR/ MWh th | 47 % | DIW (2) |
| Nuclear | 60 | - | - | 100 EUR/kW/y | 3 EUR/ MWh th | 33 % | DIW (2) |
| Run of River | - | - | - | 35 EUR/kW/y | - | - | DIW (2), IEA |
| Pumping Storage | - | - | - | 45 EUR/kW/y | - | - | DIW (2), IEA |
| Reservoir | - | - | - | 40 EUR/kW/y | - | - | DIW (2), IEA |

(1) 50Hertz Transmission GmbH, Amprion GmbH, TenneT TSO GmbH & TransnetBW GmbH. (2023). Netzentwicklungsplan Strom 2037/2045 (2023): https://www.netzentwicklungsplan.de/sites/default/files/2023-03/230321_NEP_Kostenschaetzung_NEP2037_2045_V2023_1.Entwurf.pdf

(2) : DIW (2013): Current and prospective costs of electricity generation until 2050, https://www.diw.de/documents/publikationen/73/diw_01.c.424566.de/diw_datadoc_2013-068.pdf

# APPENDIX C

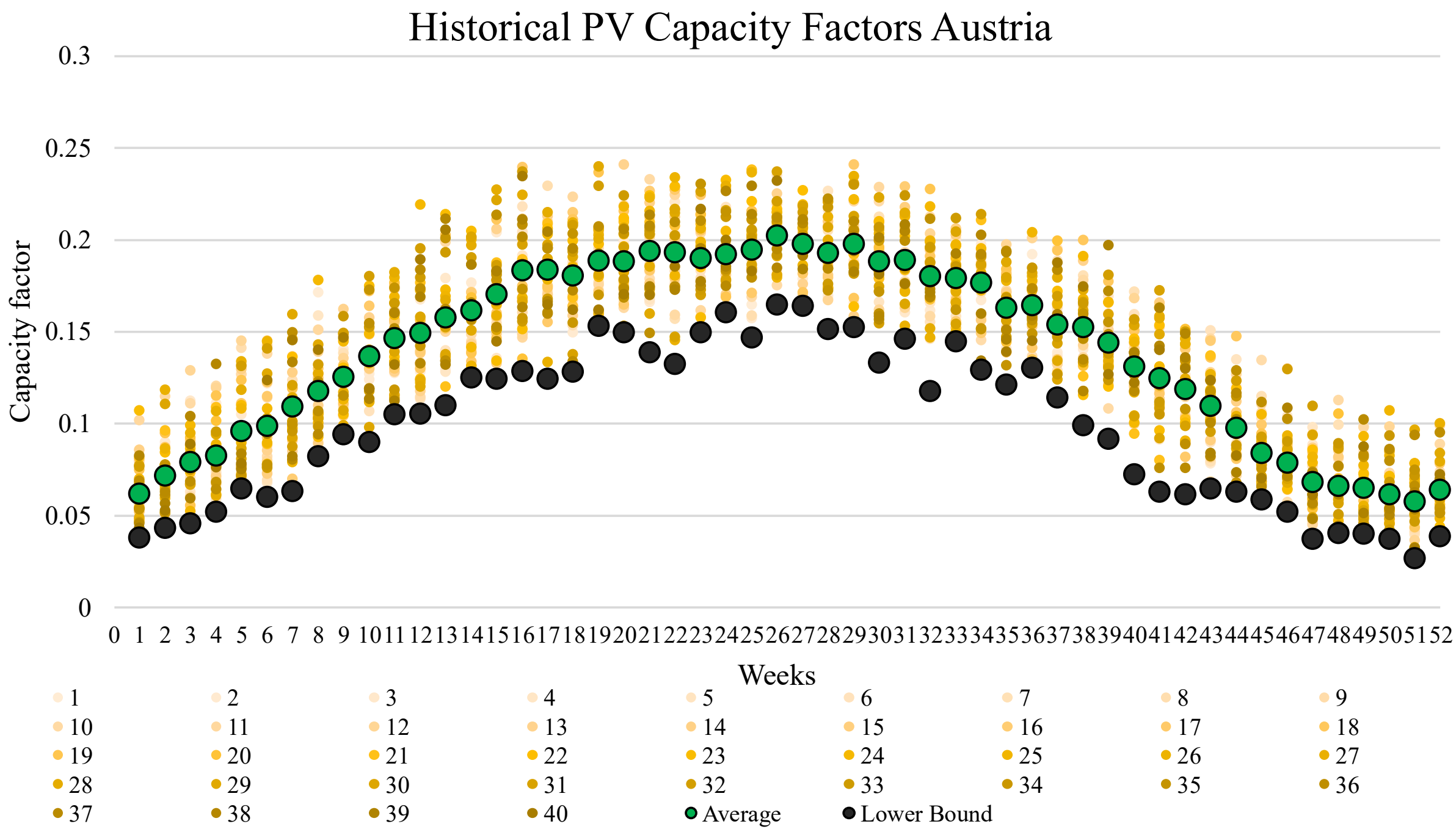

**Figure 10:** Averaged weekly solar PV capacity factors in Austria, with average weeks in green and lower bounds in black, own illustration.

# APPENDIX D

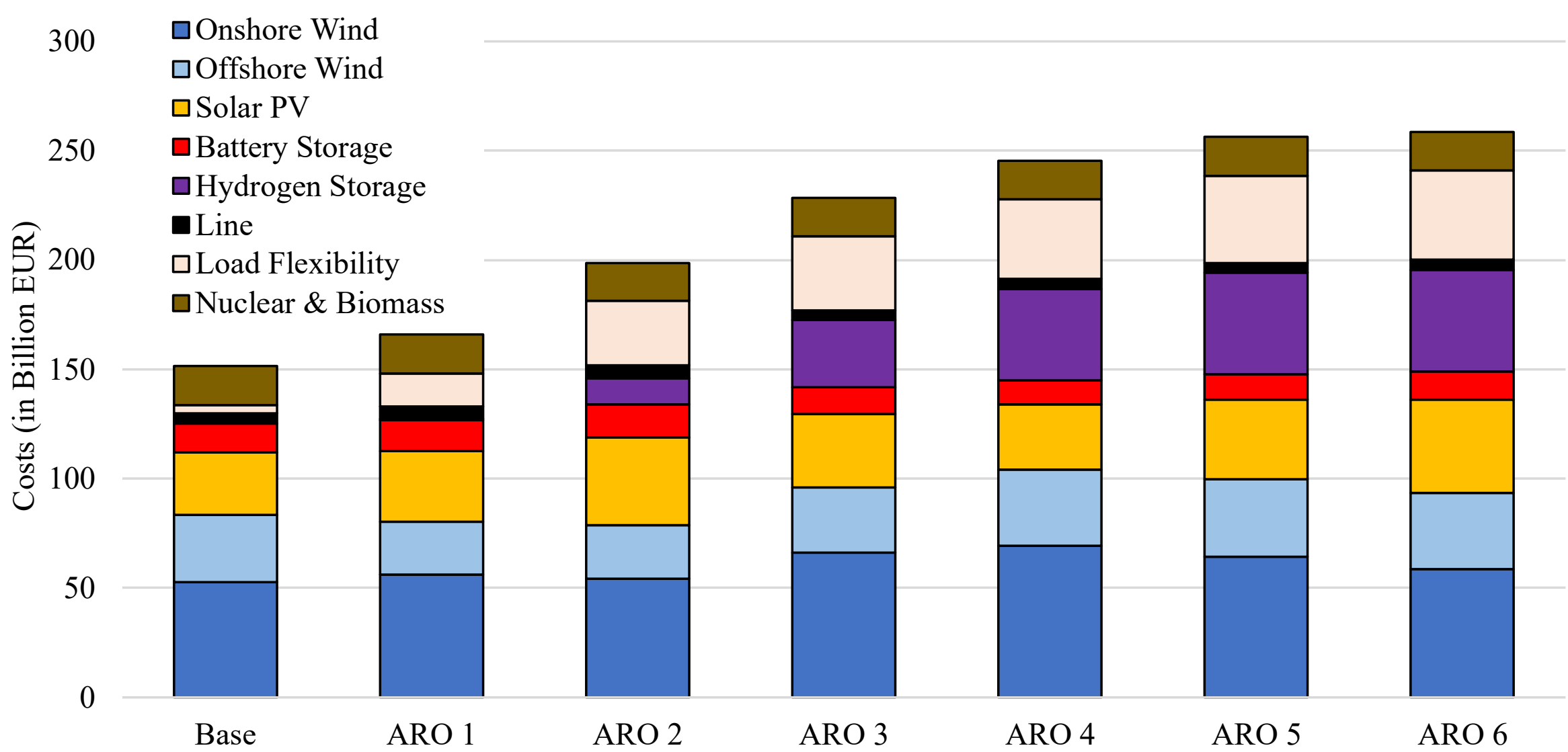

**Figure. 11.** Scenario specific investments and operation costs, own illustration.

Figure. 11 shows the costs per scenario differentiated by generation, storage and transmission technology. As we are analyzing a carbon neutral future electricity system, the majority of the costs originate from investments in renewable generation technologies. In the base scenario (left column), wind onshore is the preferred investment option, with total investments amounting to around € 50 bn. Wind offshore and solar PV follow with costs of around € 30 bn each. About € 10 bn is invested in battery storage technologies, smaller amounts in transmission line expansions and biomass or nuclear operation. The base scenario does not invest in long term hydrogen storage technologies. Increasing the uncertainty budget, i.e. moving from ARO1 to ARO6 in the graph, we can now deduce which technologies are covering the low wind and solar availabilities. Up to ARO3, the total installed capacity of RES (onshore wind, offshore wind and solar) increases. Additional capacity (MW) thus compensates for reduced availability (MWh per MW). Furthermore, investment shifts to regions which are not affected by reduced renewable availability. From ARO3 to ARO6, aggregated RES capacity remains relatively constant because increasingly more regions are affected by reduced solar or wind availability. This is because expanding or reallocating renewable capacity becomes less effective in mitigating the impact. Instead, the model prioritizes the increasing use of load shedding and invests in seasonal hydrogen storage. In contrast, the share of battery storage remains almost constant from the ARO3 scenario, because it cannot contribute to shift energy beyond the daily balancing use of residual loads. The same applies to grid expansion, although at significantly lower investment levels.

# APPENDIX E

Figure 12 below plots the endogenous expansion of transmission line capacity for each region (left axis), differentiated between the cross-regional capacity and the regional (internal) transmission capacity. Additionally, it presents the import-export balance of the individual regions during the extreme event period (right axis). The figure shows that the slightly lower total transmission capacity in the ARO3-ARO6 scenarios originates from the reduced internal transmission line expansion in Region 6. Furthermore, the figure reveals that most cross-border capacity is built in Region 3 and Region 6. This is in line with the corresponding energy exchange between these regions, where Region 3 is a clear exporter and Region 6 a clear importer during all scenarios as indicated by Figure 7. The scenario specific line expansion on a geographical scale is visualized in the supplementary materials.

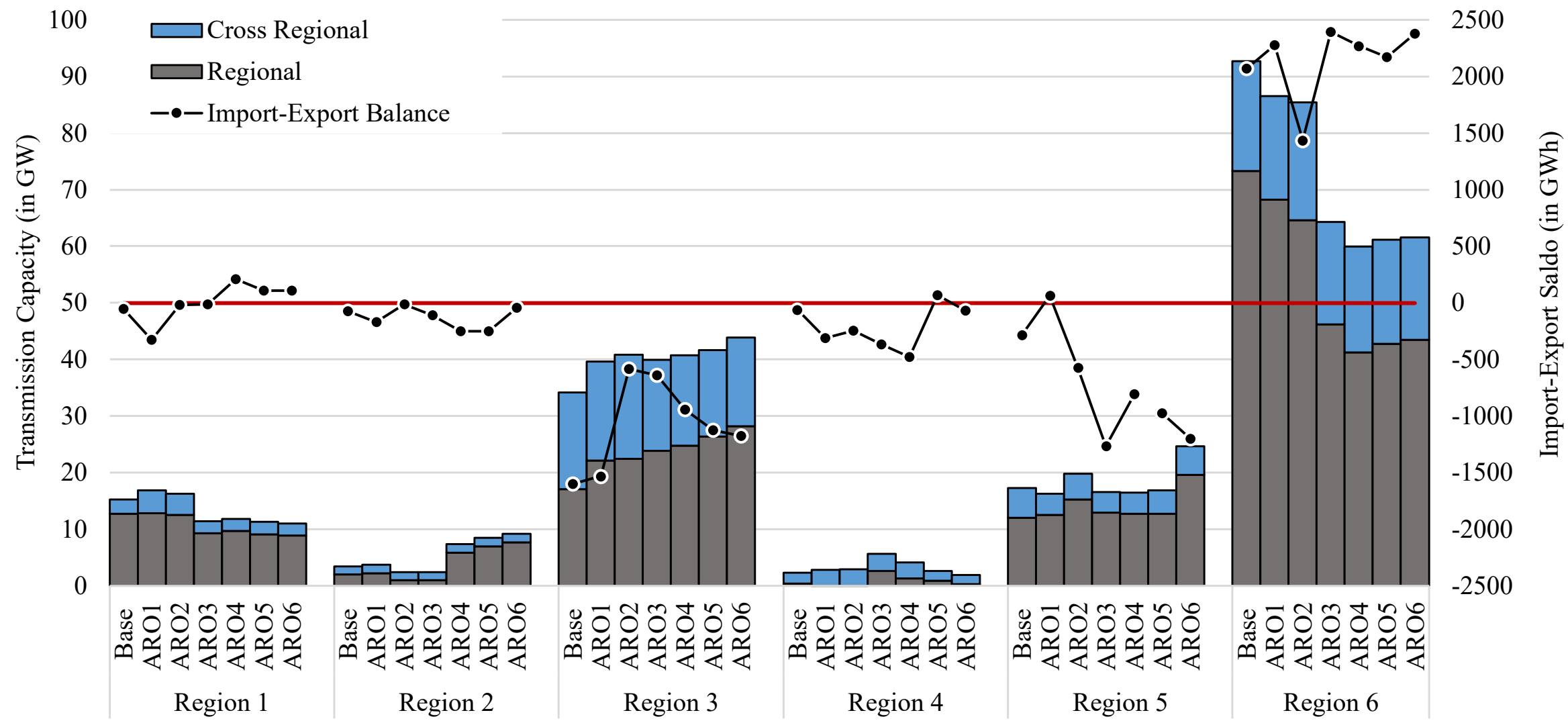

**Figure 12:** Transmission capacity expansion and import-export balance during extreme event period (January), own illustration.

The dominating importing region throughout all scenarios is Region 6. This result is driven by two factors. The renewable availability on the one hand, and the demand centers on the other hand. Region 6 has the highest electricity demand and comparably low renewable generation potentials compared to the other regions. Furthermore, the share of the firm capacity (exogenous hydro, biomass, and nuclear power), is the smallest compared to the overall demand. This is leading to a system layout, where Region 6 imports cheap renewable electricity generated in neighboring regions with a better renewable potential such as Region 1, Region 3, and Region 5.

We also find that depending on the modeled extreme event, Region 6 is influencing the capacity and generation mix in its neighboring regions. For example, while Region 3 is utilizing offshore wind in the Base scenario, it switches to onshore wind generation in the ARO scenarios. In contrast, Region 6 builds more offshore wind capacity. The system seeks to minimize the distance between generation and demand centers: onshore wind is now prioritized in Region 3 due to its proximity to Region 6, while offshore wind is favored in Region 6 because, even during low wind availability events, offshore wind has greater generation potential than onshore wind.

# APPENDIX F

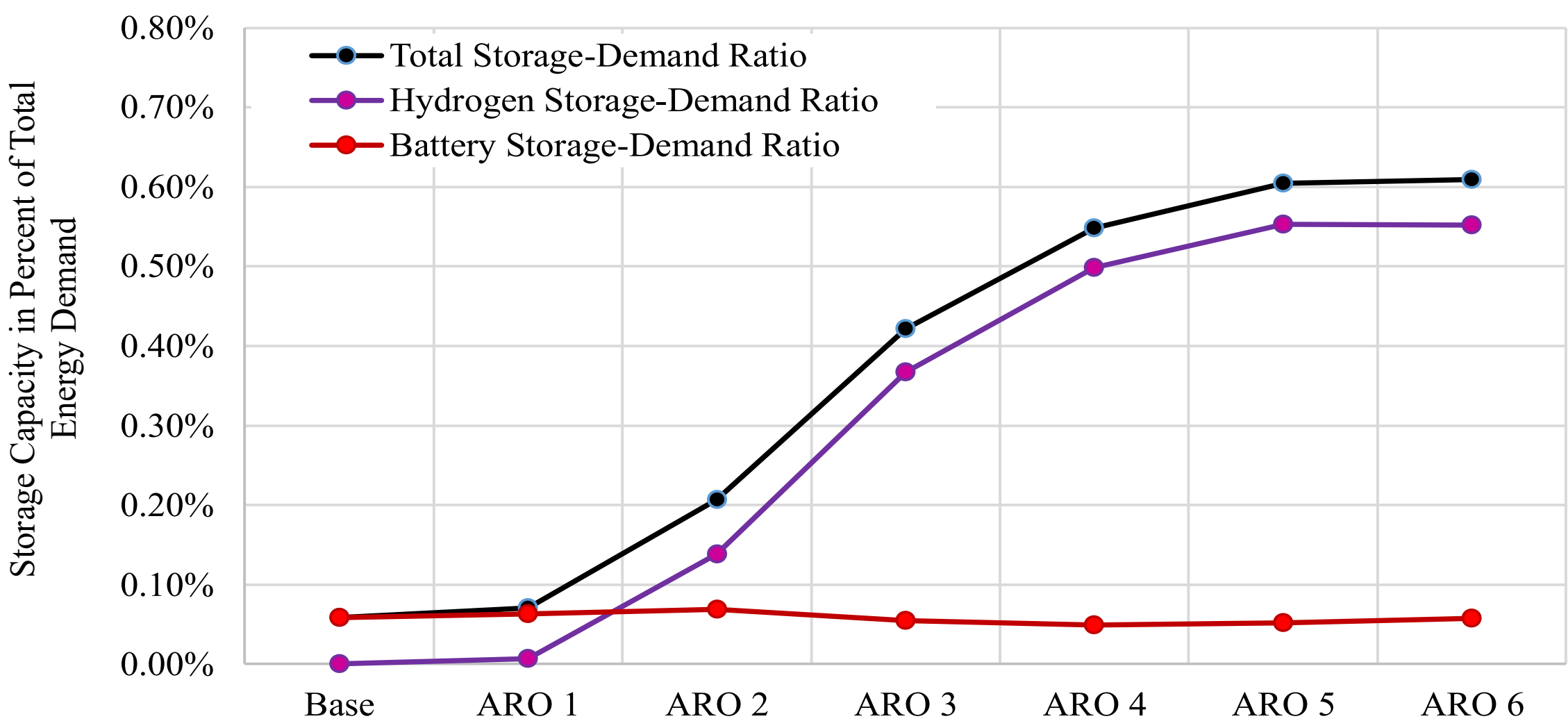

**Figure 13:** Scenario and technology specific storage capacity demand ratio, own illustration.

Figure 13 shows that the total storage-demand ratio falls between 0.08 % for the Base and 0.7 % for the ARO6 scenario. This is in line with estimates for 100% renewable systems Zerrahn et al. (2018). The ratio significantly increases from scenario ARO2 onwards. This increase is related to the expansion of long-term hydrogen storage. The hydrogen storage capacity increases strongly in the ARO2 to ARO4 scenarios compared to the Base scenario while it is the same in ARO5 and ARO6. On the contrary, the battery storage capacity remains almost constant throughout all scenarios.

The regional specific energy storage to demand is depicted in Figure 14 below. The figure shows that Region 1, Region 3 and Region 5 proportionally drive the storage expansion the most.

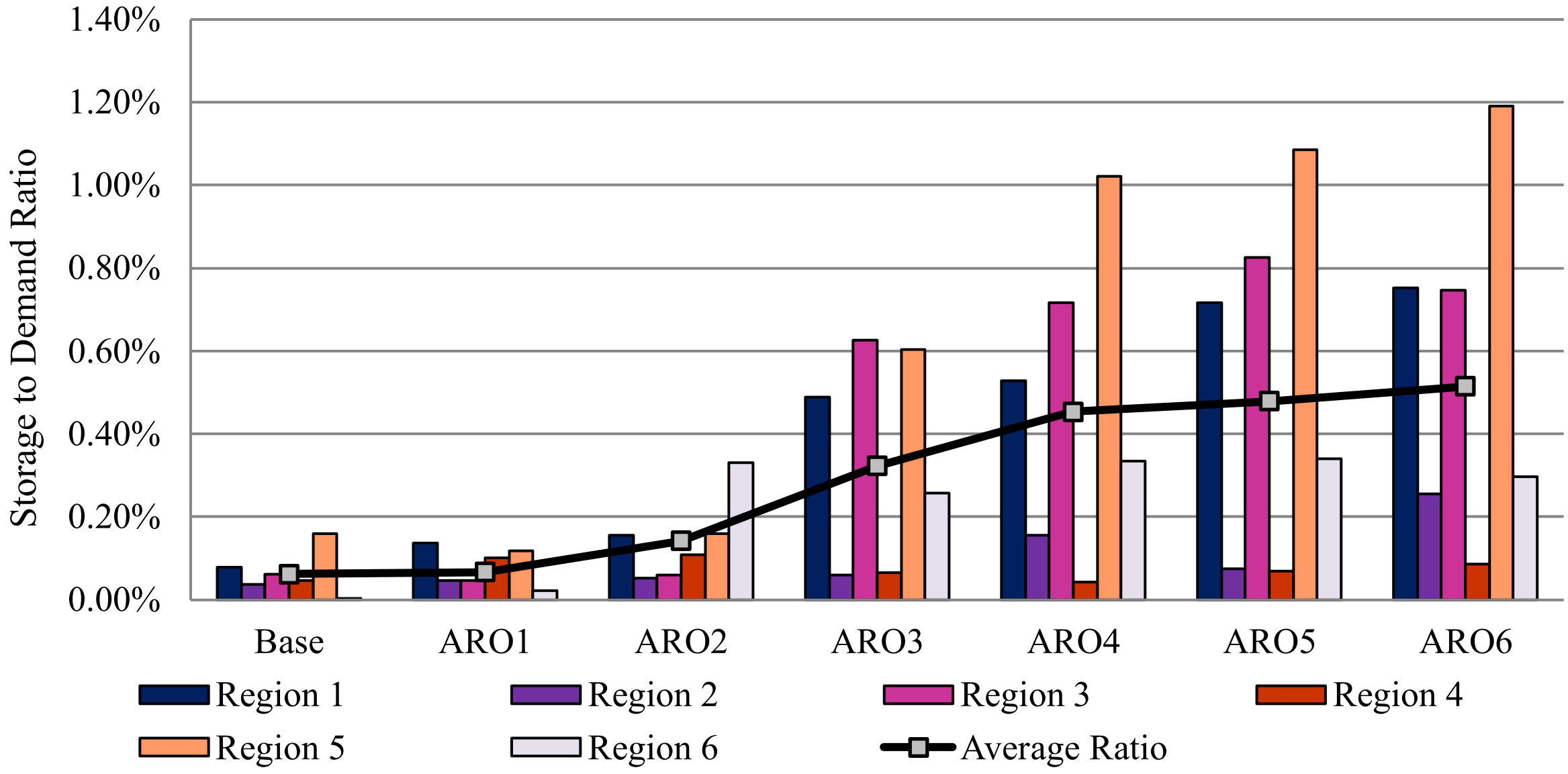

**Figure 14: Scenario specific regional comparison of storage capacity to demand ratio.**

# APPENDIX G

Figure 15 illustrates the convergence process for each ARO scenario, with the duality gap plotted on a logarithmic scale (y-axis) over the number of iterations (x-axis). The red line indicates the convergence tolerance that serves as the stopping criterion for the column and constraint algorithm.

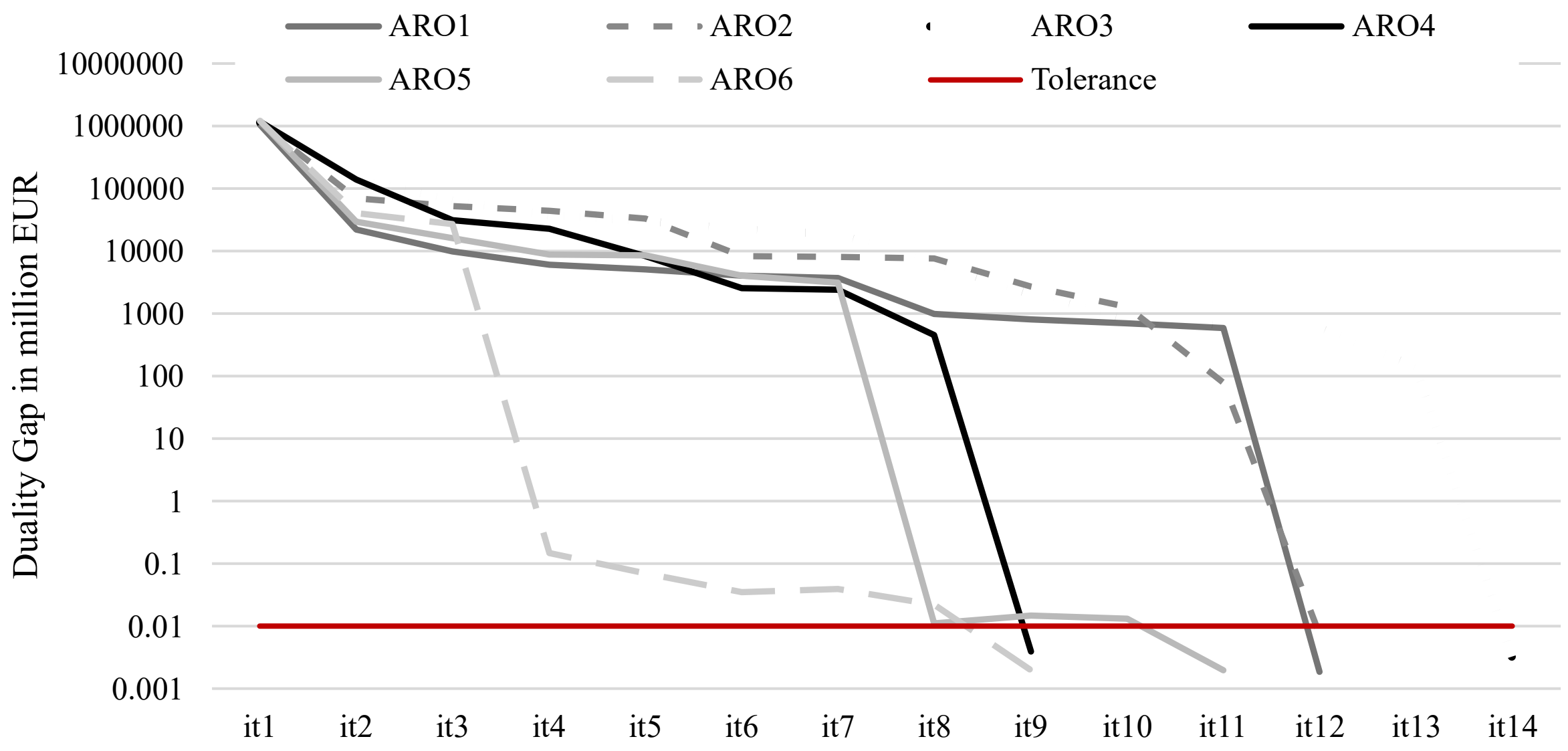

**Figure 15: Convergence Process - Development of the Duality Gap for each ARO Scenario**

The figure reveals a consistent three phase convergence progress of the ARO algorithm. In the first phase, the algorithm rapidly minimizes the duality gap up to a certain cost interval between 100 billion and 10 billion EUR. From here the second phase can be seen as saturation phase, where the min-max algorithm slowly progresses on closing the duality gap. In the third and final phase, a steep decline in the duality gap is observed, reflecting the identification of near-robust or robust capacity layout that meets the convergence criterion. As one can see, this process is different for each ARO scenario. Scenarios such as ARO1, ARO2, and ARO3 require more iterations to converge. This indicates that identifying a robust system layout in these cases requires exploring a broader set of adverse weather events before achieving convergence. The resulting extreme event realization in the last iteration is then referred to as the worst-case geographical configuration within the given uncertainty budget.